\documentclass[prb,twocolumn,longbibliography,superscriptaddress,preprintnumbers, nofootinbib, raggedbottom]{revtex4-2}
\usepackage[colorlinks,bookmarks=true,citecolor=blue,linkcolor=blue,urlcolor=blue]{hyperref}
\usepackage{dcolumn,graphicx,amsfonts,amsthm,bm,color,appendix,float}

\usepackage{braket}
\usepackage[normalem]{ulem}
\usepackage[version=4]{mhchem}
\usepackage{siunitx}
\usepackage{amsmath}
\usepackage{amssymb}
\usepackage{CJKutf8}
\usepackage{mathtools}
\usepackage{enumitem}
\usepackage{booktabs}
\usepackage{bbold}
\usepackage{subcaption}
\usepackage{caption}
\usepackage{multirow}

\renewcommand{\v}[1]{\boldsymbol{#1}}

\begin{document}
\title{
Berry curvature from plane-wave mixing: A three-band model with emergent symmetries
}

\author{Avi Vadali}
\affiliation{Department of Physics, Harvard University, Cambridge, MA 02138, USA}
\affiliation{California Institute of Technology, Pasadena, CA 91125, USA}
\affiliation{Department of Physics, Massachusetts Institute of Technology, Cambridge, MA 02139, USA}

\author{Tomohiro Soejima (\begin{CJK*}{UTF8}{bsmi}副島智大\end{CJK*})}
\affiliation{Department of Physics, Harvard University, Cambridge, MA 02138, USA}
\affiliation{Center for Computational Quantum Physics, Flatiron Institute,
162 5th Avenue, New York, NY 10010, USA}
\affiliation{Center for Quantum Phenomena, Department of Physics,
New York University, 726 Broadway, New York, New York 10003, USA}

\begin{abstract}
Two-dimensional materials can host topological phases stabilized either by an intrinsic moir\'{e} potential or topological crystallization. Crucial to the topology of these phases is the interplay between the (self-consistently generated) potential and the spinor structure, which imbues the band-projected potential with momentum-dependence. Truncating this potential at linear-order in momentum, we obtain a three-band model that captures the low-energy band structure and Berry curvature of such topological bands. We analytically find that the Berry curvature strongly depends on momentum-dependence of the potential. Furthermore, we identify that the model can be tuned to have emergent symmetries, including $U(1)$ and anti-unitary symmetries. Emergent-symmetry breaking perturbations drive the system to regimes hosting topological flat bands, quadratic band touchings, and multiple Dirac cones. We recover many of these features when studying an un-truncated model of a $\lambda$-jellium spinor inside of a periodic potential. Our results reveal a rich interplay of spinor structure and a potential in determining the topology of the ground state.
\end{abstract}

\maketitle

\section{Introduction}
\label{sec:introduction}

Material properties are often governed by the topological structure of their electronic wavefunctions. In two dimensions, the quantized Hall conductance of integer quantum Hall states and Chern insulators reflects the topological character of their Bloch bands, manifesting as chiral edge transport \cite{TKNN, quantized_hall_conductance, bulk_boundary, hasan_kane, HaldaneModel}. The topology of a band is encoded in its Berry curvature, which quantifies the momentum-space winding of the Bloch wavefunction.

Moir\'{e} materials recently emerged as a versatile platform for realizing topological electronic states \cite{bistrizer_macdonald, AHF_TBG, pseudo_LL_TBG, TI_in_TMD, magic_in_TMD, li2021quantum, zou2018band, po2018origin, po2019faithful, song2019all}.
In these systems, the lattice mismatch between stacked layers produces a potential that reconstructs the electronic bands. When combined with the spinor structure of the original electronic bands, this moir\'{e} potential can yield minibands with nonzero Berry curvature and Chern numbers. Such moir\'{e}-induced Chern bands underpin a wide class of topological materials
from twisted bilayer graphene (TBG) \cite{AHF_TBG, pseudo_LL_TBG, bistrizer_macdonald, zou2018band, po2018origin, po2019faithful, song2019all}, where the minibands inherit the topology arising from Dirac cones of monolayer graphene, to transition-metal dichalcogenide (TMD) heterostructures \cite{TI_in_TMD, magic_in_TMD, li2021quantum}, where nontrivial texture of the potential can hybridize plane waves to generate topological bands. 

The hybridization of plane waves is a general mechanism for generating topological bands in many different settings~\cite{zeng_gate_tunable, ghorashi_bg, tan_semicond, ideal_AHC, lambda_N_jellium, sublattice_structure, lambda_jellium, sc_ring_model, AHCII, berry_trashcan, FQAH_moireless, AHCI}. Breaking continuous translation symmetry to discrete translation symmetry identifies previously distinct momenta onto one another, and the resulting plane-wave mixing produces finite Berry curvature. Such translation symmetry-breaking can either be induced using an external periodic potential \cite{zeng_gate_tunable, ghorashi_bg, tan_semicond, ideal_AHC, lambda_N_jellium} or driven by interactions, as is the case in the transition from an electron gas to an anomalous Hall crystal (AHC) \cite{AHCI, AHCII, ideal_AHC, sc_ring_model, sublattice_structure, lambda_N_jellium, berry_trashcan, FQAH_moireless, wavefunction_FAHC, lambda_jellium, extended_QAH}.
Many works have analyzed the important role played by spinor structure~\cite{AHCII, lambda_N_jellium, ideal_AHC, sc_ring_model, lambda_jellium, sublattice_structure, berry_trashcan, tan_semicond}, yet a simple analytical account of how this Berry curvature arises remains lacking.

In this work, we study the generation of topology when a class of nontrivial spinors is subjected to a $C_6$-symmetric potential. By focusing on the low energy behavior near high-symmetry points, we obtain a simple three-band model parametrized by  three variables $\Delta, \alpha, v_F$. We then compute expressions for the Berry curvature, its roots, and its divergences at the origin as functions of $\Delta, \alpha, v_F$. The result is widely applicable to a large class of microscopic models.

Intriguingly, the model can be tuned to have a variety of emergent symmetries. We identify conditions for the emergence of $U(1)$ and anti-unitary symmetries with associated enforced trivial/topological flat bands. Perturbing away from these points allows us to identify rich phenomenology, including quadratic band touchings and multiple satellite Dirac cones.

To test the validity of the three-band model, we apply it to $\lambda$-jellium~\cite{lambda_jellium}, which has been shown to exhibit rich topological physics~\cite{lambda_jellium, lambda_N_jellium}. We numerically verify that the three-band model recovers the band structure and Berry curvature of the un-truncated model near topological critical points. Moreover, we observe the quadratic band touching and multiple Dirac cones predicted by the effective model within the un-truncated model.
These results demonstrate the ability of the three-band model to capture much of the band topology generated by the combination of a periodic potential and the internal structure of an entire family of spinors.

The remainder of this paper is organized as follows. Sec.~\ref{sec:effective_hamiltonian} introduces the effective 3-band model. Sec.~\ref{sec:spinorless_phenomenology} investigates the spinorless Berry curvature contribution and identifies critical points. Sec.~\ref{sec:emergent_symmetries} identifies emergent symmetries of the 3-band model.  Sec.~\ref{sec:lim_topo_flat_bands} and Sec.~\ref{sec:lim_triv_flat_bands} each discuss the band structures and Berry curvatures of the 3-band model near distinct fine-tuned points. We close by demonstrating the accuracy of the 3-band approximation when applied to $\lambda$-jellium spinors subject to a periodic potential in Sec.~\ref{sec:connection_to_full_model}.

\section{Effective Hamiltonian}
\label{sec:effective_hamiltonian}

We consider a class of Hamiltonians over a Hilbert space spanned by plane waves states $\ket{\v{k}, a} = \int d^2 \v{r} e^{i \v{k} \cdot \v{r}} \ket{\v{r}, a}$ where $\v{k}$ is the momentum and $a$ indexes the orbital component. The corresponding creation operator is given by $f^\dagger_{\v{k}, a}$, and the kinetic portion of the Hamiltonian is written in second quantization as
\begin{equation}
    \hat{H} = \sum_{\v{k}, a, b} H(\v{k})_{ab} f^\dagger_{\v{k}, a} f_{\v{k}, b} = \sum_{\v{k}, n} E_{n}(\v{k}) c^\dagger_{\v{k}, n} c_{\v{k}, n}.
\end{equation}
Single-particle eigenstates of $\hat{H}$ (indexed by $n$) are given by 
$\ket{\chi_n(\v{k})} = \sum_a \chi_n^a(\v{k}) \ket{\v{k}, a}$ with the corresponding creation operator $c_{\v{k}, n}^\dagger = \sum_a \chi_n^a(\v{k}) f_{\v{k}, a}^\dagger$. We are interested in the low energy states of $\hat{H}$, so we drop the band indices: $\ket{\chi(\v{k})}$ is the low-energy state and $c_{\v{k}}^\dagger \ket{0} = \ket{\chi(\v{k}) }$. We assume that the Hamiltonian has a $C_3$ symmetry whose action on the basis states is specified by $C_3 \ket{\v{k}, a} = \lambda^a \ket{C_3 \v{k}, a}$ where $\lambda^a$ are the $C_3$ eigenvalues. In this work, we focus on parent spinors that are holomorphic in momentum $q = q_x + i q_y$, i.e. $\frac{\partial \chi^a}{\partial \overline{q}} = 0$, and we further assume that the parent spinors are invariant under the mapping $q \rightarrow \overline{q}, \chi^a \rightarrow \overline{\chi}^a$.

A periodic potential reconstructs this band structure. We study a $C_6$ symmetric potential comprising the first harmonics which is diagonal in the orbital basis:
\begin{equation}
    \hat{V} = \int d^2 \v{r} \sum_a V^a \ket{\v{r}, a}\bra{\v{r}, a}\sum_j \cos(\v{g}_j \cdot \v{r}).
\end{equation}
Here the reciprocal lattice vectors are $\v{g}_1 = \frac{2 \pi}{a \sqrt{3}} (\sqrt{3}, -1)^T, \v{g}_2 = \frac{4 \pi}{a \sqrt{3}} (0,1)^T, \v{g}_3 = -\v{g}_1 - \v{g}_2$ with lattice spacing $a$.

\par Projected to the lowest band, the potential can be written in second quantization as
\begin{equation}
\label{eq:moire_pot}
    \begin{split}
        \hat{V} &= \sum_{\v{k}, j, a} \frac{V^a}{2} f^\dagger_{\v{k}+ \v{g}_j, a} f_{\v{k}, a} + \mathrm{h.c.} \\
        &= \sum_{\v{k}, a, j} \frac{V^a}{2} \overline{\chi^a(\v{k} + \v{g}_j)} \chi^a(\v{k}) c^\dagger_{\v{k} + \v{g}_j} c_{\v{k}} + \mathrm{h.c.} \\
        &= \sum_{\v{k}, a, j} V^a_j (\v{k}) c^\dagger_{\v{k} + \v{g}_j} c_{\v{k}} + \mathrm{h.c.}
    \end{split}
\end{equation}
where we introduced
\begin{equation}
    \label{eq:V_non_uniform}
    \begin{split}
        V^a_j (\v{k}) &= \frac{V^a}{2} \overline{\chi^a(\v{k}+ \v{g}_j)} \chi^a(\v{k}).
    \end{split}
\end{equation}
Since this matrix element is not gauge invariant under $\chi^a(\v{k}) \to e^{i\theta(k)}\chi^a(\v{k})$, all expressions below are understood within a fixed gauge adopted for $\chi^a$.

\begin{figure}
    \centering
    \includegraphics[width=0.3\textwidth]{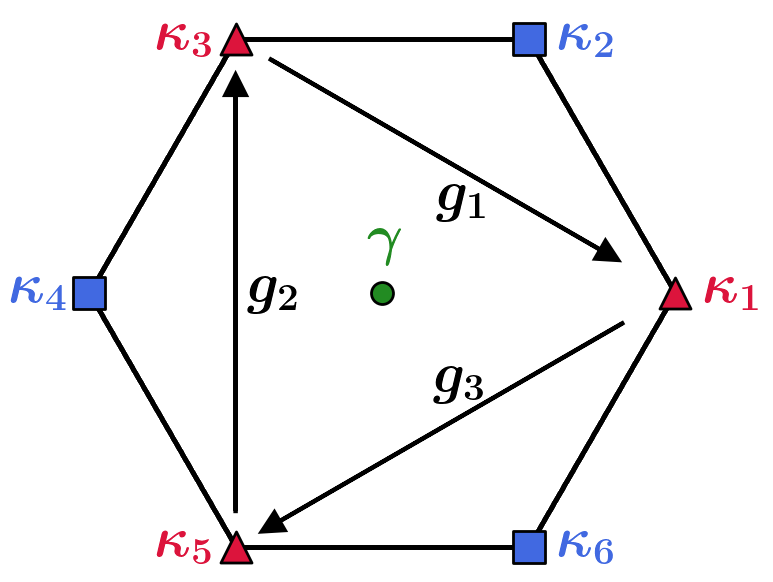}
    \caption{Mini-Brillouin zone high-symmetry points. Here $\v{\kappa}_{2n + 1}$ span the $\v{\kappa}$ point and $\v{\kappa}_{2n}$ span the $\v{\kappa}'$ point. The gamma point lies at the center of the BZ. The reciprocal lattice vectors are denoted by $\v{g}_j$.
    }
    \label{fig:mBZ_labels}
\end{figure}

\subsection{Three-band Hamiltonian}
\label{sec:3band_ham}

The periodic potential induces band folding into a Brillouin zone (BZ), which contains three high-symmetry points that are invariant under the action of $C_3$: $\v{\gamma} \equiv (0, 0)$, $\v{\kappa} \equiv (\v{g}_1 - \v{g}_3) / 3$, and $\v{\kappa}' \equiv (\v{g}_3 - \v{g}_1) / 3$ (Fig.~\ref{fig:mBZ_labels}). We focus on the $\v{\kappa}$ point in this work. The momenta with smallest magnitude that get folded to $\v{\kappa}$ are given by $\v\kappa_1 = \frac{\v{g}_1 - \v{g}_3}{3}, \v\kappa_3 = \frac{\v{g}_2 - \v{g}_1}{3}, $and $\v\kappa_5 = \frac{\v{g}_3 - \v{g}_2}{3}$. In the presence of $C_3$ symmetry, these momenta have degenerate energies i.e., $E(\v\kappa_1) = E(\v\kappa_3) = E(\v\kappa_5)$.

To capture the low-energy behavior around $\v{\kappa}$ point, we expand the kinetic term $T$ and the potential $V$ to linear order.
We express the resulting three-band Hamiltonian in the basis spanned by $\{\ket{\chi(\v{\kappa}_n + \v{q})} \}$ with creation operators $c_{\v{q}, n}^\dagger = \sum_a \chi^a(\v{\kappa}_n + \v{q}) f_{\v{\kappa}_n + \v{q}, a}^\dagger$. Expanding the dispersion to linear order around $\v{\kappa}$ yields 
\begin{equation}
\begin{split}
    \hat{T} &= v_F \sum_{\v{q}, n} \v{q} \cdot \hat{\v{e}}_n c^\dagger_{\v{q}, n} c_{\v{q}, n} \\
    &= \sum_{\v{q}, n}T_n(\v{q}) c^\dagger_{\v{q}, n} c_{\v{q}, n},
\end{split}
\end{equation}
where $\hat{\v{e}}_n$ is the unit vector parallel to $\v{\kappa}_n$ and $v_F$ is the signed velocity at the high-symmetry point . For convenience, we rewrite the linearized dispersion in terms of the (anti-)holomorphic momenta $\overline{q} = q_x - i q_y, q = q_x + i q_y$ as
\begin{equation}
\label{eq:T_n_def}
\begin{split}
    T_n(\v{q}) = \frac{v_F}{2} (\omega^{2 +n} q + \overline{\omega}^{2 + n} \overline{q}) ,
\end{split}
\end{equation}
where $\omega = \exp(i 2 \pi/3)$ and $n \in \{1, 3, 5\}$. 

From Eq.~\eqref{eq:moire_pot}, the potential is
\begin{equation}
\label{eq:V_pot_def}
    \hat{V} = \sum_{\v{q}, a, m, n} V^a_{m, n}(\v{q}) c_{\v{q}, m}^\dagger c_{\v{q}, n},
\end{equation}
where the potential matrix elements are
\begin{equation}
    \label{eq:V_mat_elems}
    V^a_{m, n}(\v{q}) = \frac{V^a}{2} \overline{\chi^a(\v{\kappa}_m + \v{q})} \chi^a(\v{\kappa}_n + \v{q}).
\end{equation}
We are interested in the case $m \neq n$, so we re-parameterize these matrix elements in terms of a single index as $V_1^a(\v{q}) \equiv V_{1, 3}^a(\v{q}), V_3^a(\v{q}) \equiv V_{3, 5}^a(\v{q}), V_5^a(\v{q}) \equiv V_{5, 1}^a(\v{q})$. 

The linear-in-$\v{q}$ Taylor expansion of these matrix elements is constrained by two key symmetries. First, the potential terms satisfy the  $C_3$ relation
\begin{equation}
\label{eq:c3_pot_relation}
    V^a_{1}(\v{q}) = V^a_{3}(C_3 \v{q}) = V^a_{5}(C_3^2 \v{q}).
\end{equation}
Second, the invariance of $\chi^a(k)$ under $k \rightarrow \overline{k}$ implies $\chi^a(\v{\kappa}_j + \v{q})$ is invariant under the mapping $\v{\kappa}_j + \v{q} = (\kappa_{j,x} + q_x, \kappa_{j,y} + q_y) \rightarrow (\kappa_{j,x} + q_x, -\kappa_{j,y} - q_y)$. Since $(\kappa_{3, x}, \kappa_{3, y}) = (\kappa_{5, x}, -\kappa_{5, y})$, the aforementioned symmetry reveals that $V_3(\v{q})$ is preserved by $q_y \rightarrow -q_y$. 

Considering this inversion symmetry with the $C_3$ relation in Eq.~\eqref{eq:c3_pot_relation}, the linear-order Taylor expansion of $V_m(\v{q})$ is restricted to the form
\begin{equation}
\label{eq:holo_ff}
    V_{m}(\v{q}) \approx W_m(\v{q}) \equiv \Delta + \frac{\alpha}{2} \bigg(\omega^{m} q + \overline{\omega}^{m} \overline{q} \bigg)
\end{equation}
for $m \in \{1, 3, 5\}$. We define the truncated potential $\hat{W}$ in analogy to $\hat{V}$ in Eq.~\eqref{eq:V_pot_def}. Here the zeroth order term is 
\begin{equation}
    \Delta = V_{1}(\v{q}=0) = V_{3}(\v{q}=0) = V_{5}(\v{q}=0),
\end{equation}
and the linear order coefficient is
\begin{equation}
\label{eq:ff_coeffs}
    \alpha = \bigg(\frac{\partial V_{3}}{\partial q}  + \frac{\partial V_{3}}{\partial {\overline{q}}} \bigg) \bigg|_{q=0} = \frac{\partial V_{3}}{\partial q_x} \bigg|_{q=0}.
\end{equation}

The truncated Hamiltonian, $H = T + W$, has matrix elements that can be expressed as a sum of $v_F, \Delta, \alpha$ contributions respectively as
\begin{equation}
\label{eq:ham_3x3_expression}
\begin{split}
    H(\v{q}) &= \frac{v_F}{2} \begin{pmatrix}
        q + \overline{q} & 0 & 0 \\
        0 & \overline{\omega} q + \omega \overline{q} & 0 \\
        0 & 0 & \omega q + \overline{\omega} \overline{q}
    \end{pmatrix}
    +
    \begin{pmatrix}
        0 & \Delta & \overline{\Delta} \\
        \overline{\Delta} & 0 & \Delta \\
        \Delta & \overline{\Delta} & 0
    \end{pmatrix} \\
    &+ \frac{1}{2} \begin{pmatrix}
        0 & \alpha [\omega q + \overline{\omega} \overline{q}] & \overline{\alpha} [\overline{\omega} q + \omega \overline{q}] \\
        \overline{\alpha} [\omega q + \overline{\omega} \overline{q}] & 0 & \alpha [q + \overline{q}] \\
        \alpha [\overline{\omega} q + \omega \overline{q}] & \overline{\alpha} [q + \overline{q}] & 0
    \end{pmatrix}.
\end{split}
\end{equation}
The Hamiltonian is similar to the $3 \times 3$ Hamiltonian studied in Ref.~\cite{SFB_wang_yao, sutherland_localization}.

\subsection{$\v{q}=0$ eigenproblem and symmetry indicators}
\label{sec:symm_indic}
At the point $\v{q} = 0$, the Hamiltonian reduces to the circulant form
\begin{equation}
    H(\v{q}=0)= \begin{pmatrix}
        0 & \Delta & \overline{\Delta} \\
        \overline{\Delta} & 0 & \Delta \\
        \Delta & \overline{\Delta} & 0
    \end{pmatrix}.
\end{equation}
As a $3 \times 3$ traceless circulant matrix, $H$ has eigenvalues and eigenvectors $E_\ell = 2 \text{Re}(\omega^\ell \Delta), \v{v}_\ell = (1, \omega^\ell, \overline{\omega}^\ell)$. The phase of $\Delta$ determines the ground state. 
The ground states are degenerate at $\arg \Delta = 0, \pm \frac{2\pi}{3}$, which corresponds to band inversion.

The ground state angular momentum $\ell$ at high symmetry points constrains the total Chern number as $C \bmod 3 \equiv \ell_\kappa + \ell_{\kappa'} + \ell_\gamma$ according to the theory of symmetry indicators~\cite{fu_topological_2007,turner_quantized_2012,fang_bulk_2012,po_symmetry-based_2017,bradlyn_topological_2017,po_symmetry_2020,cano_band_2021} . Therefore, the phase of $\Delta$ contains important topological information about the system~\cite{AHCII, sc_ring_model, efficient_prediction, berry_trashcan, lambda_N_jellium}.

\subsection{Spectrum}
\label{sec:spectrum}
At general $\v{q}$, the characteristic polynomial of $H$ is a depressed cubic. Using Hermiticity of $H$ in addition, the three eigenvalues of $H$ can be expressed in terms of trigonometric functions \cite{Zucker_2008} as
\begin{equation}
\label{eq:eigenvals}
    \epsilon_s = 2 \sqrt{\frac{B}{3}} \ \text{cos} \left(\frac{1}{3} \ \text{cos}^{-1} \left[ \frac{3 C}{2 B} \sqrt{\frac{3}{B}} \right] - \frac{2 \pi}{3} s \right),
\end{equation}
where $s \in \{-1, 0, 1\}$ indexes the three roots. In Eq.~\eqref{eq:eigenvals}, the terms $B, C$ are defined as
\begin{equation}
\label{eq:B_C_defs}
    \begin{split}
        B &= \frac{3}{2} (|\alpha|^2 + \frac{v_F^2}{2}) (q_x^2 + q_y^2) + 3 |\Delta|^2 \\
        C &= \frac{1}{4} q_x (q_x^2 - 3 q_y^2) (v_F^3 - 3 v_F |\alpha|^2 + 2 \text{Re}[\alpha^3]) \\
        &- \frac{3}{2}(q_x^2 + q_y^2) \text{Re}(\Delta \alpha^2 + 2 v_F \alpha \overline{\Delta}) + 2 \text{Re}(\Delta^3) .
    \end{split}
\end{equation}
Since $H$ commutes with $C_3$, its eigenvalues are invariant scalars, $\epsilon_s(\v{q}) = \epsilon_s(C_3 \v{q})$. At the $C_3$-invariant point $\v{q} = 0$, each band is a $C_3$ eigenstate with a definite angular momentum, which we use as its label. Band $\ell$ is the eigenvector $\v{u}_\ell = (1, \omega^\ell, \overline{\omega}^\ell)$, with $C_3$ angular momentum $\ell$ and energy $E_\ell = 2\,\text{Re}(\omega^\ell \Delta)$. For a given set of parameters $\Delta, \alpha, v_F$, $\epsilon_\ell(\v{q})$ denotes the root of Eq.~\eqref{eq:eigenvals} that reduces to $E_\ell$ at $\v{q} = 0$, extended to finite $\v{q}$ by continuity.

\subsection{Berry curvature decomposition}

In this section, we show how the total Berry curvature of eigenstates of the Hamiltonian can be separated into three distinct contributions.
\par We denote the eigenstates of $H$ as

\begin{equation}
\ket{\v{u}_n (\v{q})} = \sum_{j \in \{1, 3, 5\}} u_n^j(\v{q}) | \chi(\v{\kappa}_j + \v{q} )\rangle,
\end{equation}
where $n$ indexes the eigenstates by energy. Berry curvature is computed in the usual way:
\begin{equation}
\label{eq:tot_bc_def}
\begin{split}
    \Omega^\text{tot}_n (\v{q}) &= \v{\nabla} \times i \bra{ \v{u}_n(\v{q})} \v{\nabla}_{\v{q}} \ket{ \v{u}_n(\v{q}) } \\
    &= \Omega^\text{par}_n(\v{q}) + \Omega^\times_n(\v{q}) + \widetilde{\Omega}_n(\v{q}).
\end{split}
\end{equation}
In Eq.~\eqref{eq:tot_bc_def} we have defined the total Berry curvature as the sum of 3 terms. The first term appearing in Eq.~\eqref{eq:tot_bc_def} is the \textit{parent contribution}:
\begin{equation}
    \label{eq:parent_contrib_def}
    \Omega^\text{par}_n (\v{q}) = \sum_{j \in \{1, 3, 5\}} |u_n^j|^2 \Omega_\text{orig}(\v{\kappa}_j+ \v{q}).
\end{equation}
This term is a weighted sum of the parent Berry curvature $\Omega_\text{orig}(\v{k})= i\sum_a \v{\nabla} \times \overline{\chi}^a(\v{k})  \v{\nabla} \chi^a(\v{k})$. 

The second term in Eq.~\eqref{eq:tot_bc_def} is the \textit{cross term} between the parent wavefunction and moir\'{e}-generated wavefunction:
\begin{equation}
\label{eq:cross_term_def}
    \Omega_n^\times (\v{q}) = i \sum_{j \in \{1, 3, 5 \}} \v{\nabla} |u_n^j|^2 \times \langle \chi(\v{\kappa}_j + \v{q})| \v{\nabla} | \chi(\v{\kappa}_j + \v{q}) \rangle.
\end{equation}
Observe that $\Omega_n^\times$ vanishes if either the parent spinors carry trivial Berry connection or the mini-bands form a constant-weight superposition of the parent spinors. 

\par Finally, the term $\widetilde{\Omega}_n(\v{q})$ is the \textit{spinorless Berry curvature}, which we analytically study in Sec.~\ref{sec:spinorless_phenomenology}. This Berry curvature captures the contribution from the overlap between plane waves:
\begin{equation}
    \label{eq:spinorless_bc_def}
    \widetilde{\Omega}_n(\v{q}) = i \sum_{j \in \{1, 3, 5 \}} \v{\nabla} \times \overline{u}_n^j(\v{q}) \v{\nabla}u_n^j(\v{q}).
\end{equation}
We focus on the spinorless term for unconstrained $\Delta, \alpha$ due to its independence from the choice of parent spinor and its ability to approximate the spinorful Berry curvature in certain parameter regimes (see Sec.~\ref{sec:bc_scaling}).

\section{Spinorless Phenomenology}
\label{sec:spinorless_phenomenology}

Having isolated the spinorless contribution $\widetilde{\Omega}_n(\v{q})$ in Eq.~\eqref{eq:spinorless_bc_def}, we now study its phenomenology as a function of the model parameters $\Delta, \alpha, v_F$. We first analyze the zero-momentum limit (Sec.~\ref{sec:spinorless_zeromomentum}), where a closed-form expression for $\widetilde{\Omega}_\ell(0, 0)$ lets us determine its roots and map the sign of the Berry curvature across the $(\arg\Delta, \arg\alpha)$ plane for various regimes of $v_F / |\alpha|$. We then turn to finite momentum (Sec.~\ref{sec:spinorless_finitemomentum}), where in the limit $v_F/|\alpha| \gg 1$ the Berry curvature concentrates into $C_3$-related hotspots whose sign we characterize in terms of the relative phase of $\Delta$ and $\alpha$.

\subsection{Sign reversal at zero momentum}
\label{sec:spinorless_zeromomentum}
The spinorless Berry curvature demonstrates a rich phenomenology, including reversal of the sign of Berry curvature.
To obtain a simple point of analytic understanding, we determine the spinorless Berry curvature $\widetilde{\Omega}$ of each band at $\v{q} = 0$ (Appendix~\ref{app:spinorless_bc_deriv}), and we solve for the roots of the Berry curvature as functions of $v_F, \Delta, \alpha$ (Appendix~\ref{app:spinorless_bc_roots_deriv}). 

There are two distinct (but intersecting) branches of roots of $\widetilde{\Omega}_\ell(0, 0)$. The first branch of roots  appear when $\phi = \text{arg}(\Delta)$, $\theta = \text{arg}(\alpha)$ satisfy the following relation:
\begin{equation}
\label{eq:bc_roots_branch1}
\begin{split}
    \theta_\ell &= \phi_\ell + \cos^{-1} \bigg( \frac{v_F}{|\alpha|} \cos(\phi_\ell + \frac{2 \pi}{3} \ell) \bigg).
\end{split}
\end{equation}
Here $(\phi_\ell, \theta_\ell)$ denotes the roots $(\phi, \theta)$ for band $\ell$. The second branch of solutions is more subtle and given by
\begin{equation}
\label{eq:bc_roots_branch2}
    \begin{split}
        \phi_\ell &= n \pi + \tan^{-1} \bigg(\frac{3 \sin(\theta_\ell + \frac{2\pi \ell}{3})}{\frac{v_F}{|\alpha|} - \cos(\theta_\ell + \frac{2\pi \ell}{3})} \bigg) + \frac{\pi}{3} \ell
    \end{split}
\end{equation}
for $n \in \mathbb{Z}$.

\begin{figure*}
    \centering
    \includegraphics[width=1\textwidth]{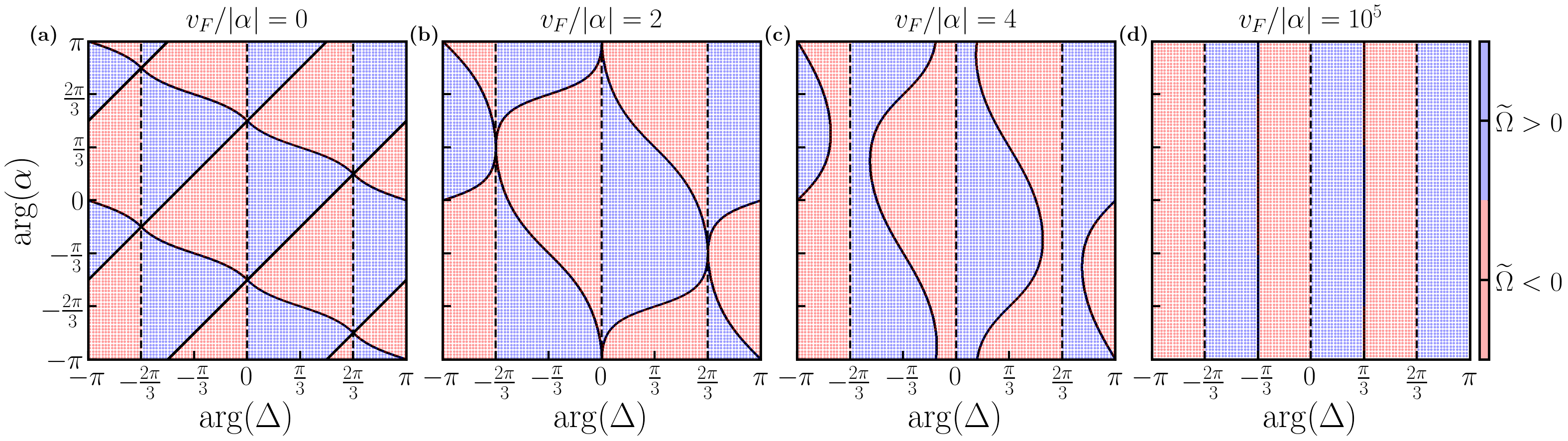}
    \caption{Sign of spinorless Berry curvature at $(q_x, q_y) = (0, 0)$ corresponding to lowest-energy band as a function of $\arg(\Delta)$ and $\arg(\alpha)$. (a-d) shows Berry curvature signs for $v_F / |\alpha| = 0, 2, 4, 10^5$. Red denotes negative Berry curvature while blue shows positive Berry curvature. The boundary (see solid black lines) where Berry curvature vanishes is analytically determined in Eq.~\eqref{eq:bc_roots_branch1} and Eq.~\eqref{eq:bc_roots_branch2}. The band crossings (see dashed black lines) produce singular behavior in $\widetilde{\Omega}_\ell(0, 0)$.
    }
    \label{fig:spinorless_ground_bc_signs}
\end{figure*}

In Fig.~\ref{fig:spinorless_ground_bc_signs}, we show the sign of $\widetilde{\Omega}(0, 0)$ for the lowest band as a function of $\Delta, \alpha, v_F$. The sign always changes at $\arg(\Delta) = 0, 2\pi/3, -2\pi/3$ due to band inversion at the origin. In addition, at small $v_F / |\alpha|$, the roots of $\widetilde{\Omega}$ forms a nearly triangular lattice (Fig.~\ref{fig:spinorless_ground_bc_signs}(a)), showing the change in $\arg(\alpha)$ has a strong impact on the Berry curvature. The intersection of the roots of $\widetilde{\Omega}(0, 0)$ and band inversion in fact signal the presence of a quadratic band touching (Sec.~\ref{sec:QBT_U(1)_PK}). The triangular features get deformed at larger $v_F / |\alpha|$, and they become stripe-like at $v_F / |\alpha|=2$. The behavior at the critical point $v_F / |\alpha| = 2$ is identified in Sec.~\ref{sec:lim_topo_flat_bands}. Finally, in the limit $v_F / |\alpha| \gg 1$ shown in Fig.~\ref{fig:spinorless_ground_bc_signs}(d), the sign distribution forms ``stripes'' that are $2\pi/3$-periodic in $\phi$. The stripes are bounded by sign changes due to band-inversion at $\phi = 0, \pm 2\pi / 3$, as well as 
continuous sign changes at $\phi = \pi, \pm \pi /3$. Thus in the large $v_F$ regime, the spinorless Berry curvature is primarily governed by $\arg(\Delta)$, as opposed to the small $v_F$ limit in which $\arg(\alpha)$ plays an important role.

\subsection{Finite momentum}
\label{sec:spinorless_finitemomentum}
At finite momentum $\v{q}$, the Berry curvature exhibits additional structure that becomes analytically tractable in the limit $v_F / |\alpha| \gg 1$. In this regime, we characterize $\widetilde{\Omega}(\v{q})$ in terms of the finite momentum scale $\Lambda_\mathrm{IR} =  4 |\Delta|/v_F$. 

For $|q| \ll \Lambda_\mathrm{IR}$, $\widetilde{\Omega}(\v{q})$ is controlled by $\Delta$ since $|\alpha| |q| \ll v_F |q| \ll |\Delta|$. As a result, $\widetilde{\Omega}(\v{q})$ is approximately rotationally-invariant, and its sign is given by the stripe distribution in Fig.~\ref{fig:spinorless_ground_bc_signs}(d).

\begin{figure}
    \centering
    \includegraphics[width=\linewidth]{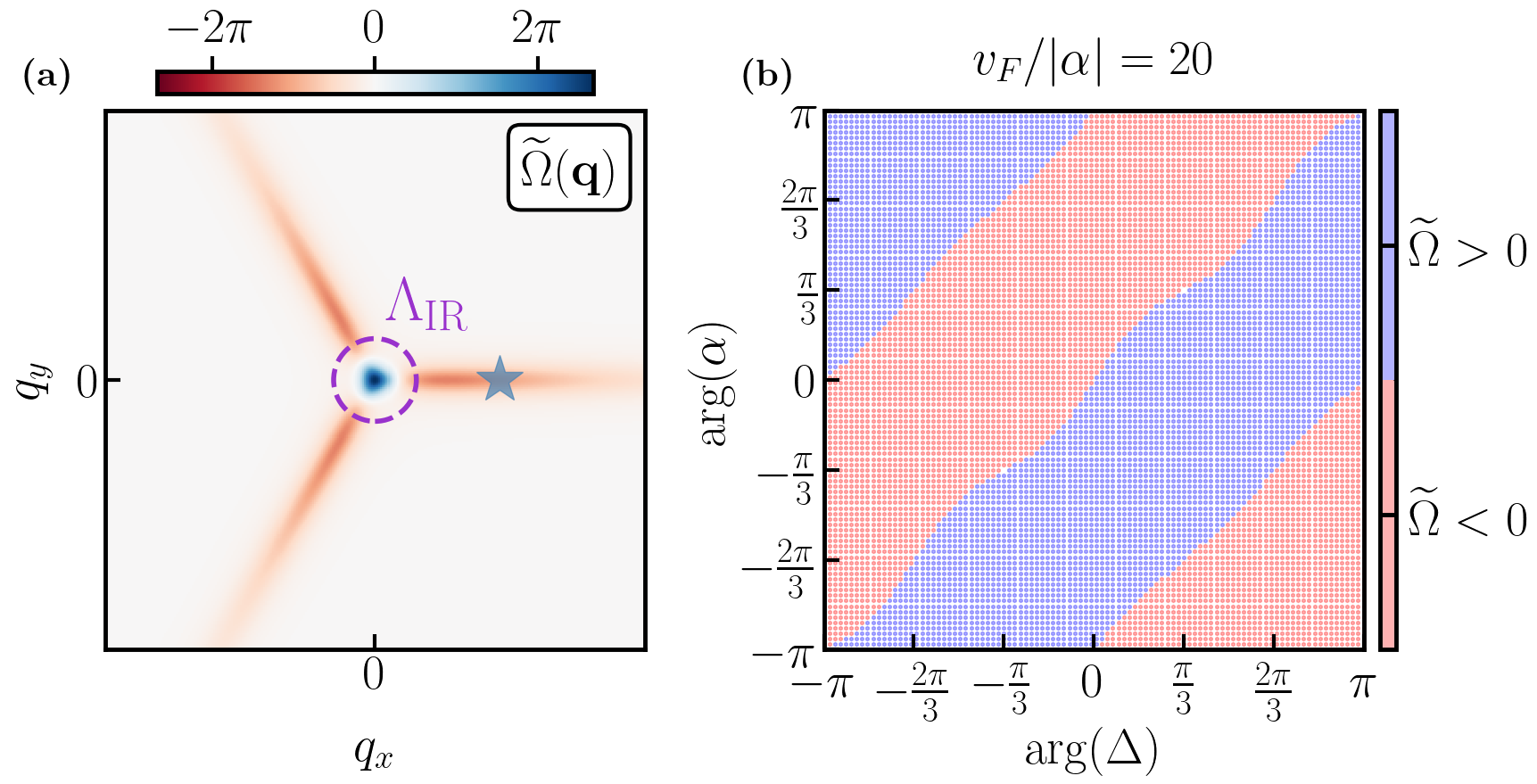}
    \caption{Spinorless Berry curvature for $v_F / |\alpha| = 20$. (a) shows $\widetilde{\Omega}$ for $\Delta = e^{i \pi/4}, \alpha = \omega$. Gray star resides at $(q_x, q_y) = (\frac{1}{2} [\Lambda_\mathrm{IR} + 1], 0)$. (b) shows sign of $\widetilde{\Omega}$ at the gray star.}
    \label{fig:large_vF_berry_distrib}
\end{figure}

At larger $|q|$,  $v_F |q|$ prefers band crossings along $\arg(q) = 0, \pm \frac{2 \pi}{3}$ while finite $\Delta$ and $\alpha$ induce a gap, producing Berry curvature along the $C_3$ related lines in Fig.~\ref{fig:large_vF_berry_distrib}(a). The sign of the Berry curvature hotspots is approximately determined by the relative phase of $\Delta$ and $\alpha$, as shown in Fig.~\ref{fig:large_vF_berry_distrib}(b). We can analytically identify changes in the hotspot sign by expanding the (un-normalized) eigenvectors of $H$ to 2\textsuperscript{nd} order in $|\Delta|/(v_F |q|), |\alpha|/v_F$ as
\begin{equation}
\label{eq:expanded_u}
    \v{u}_\ell \approx  
    \begin{pmatrix}
        [2T_{j} - T_3 - T_5]f_\ell + [T_j - T_3] [T_j - T_5] - |W_{3}|^2 \\
        \overline{W}_{1}(T_{j} - T_5) + W_{3} W_{5} \\
        W_5 (T_{j} - T_3) + \overline{W}_{1} \overline{W}_{3}
    \end{pmatrix},
\end{equation}
where $T_j$ ($j \in \{1,3,5\}$) is the kinetic term dominating band $\ell$ at large $|q|$ and $f_{\ell} \in \mathbb{R}$ is the subleading correction to $\epsilon_\ell$. A derivation of this expansion is given in Appendix~\ref{app:berry_scales}. The leading-order contributions to Berry curvature come from $\overline{W}_{1}(T_{j} - T_5)$ and $W_5 (T_{j} - T_3)$, whose phases are fixed by $\overline{W}_{1}$ and $W_5$ respectively (since $T_j \in \mathbb{R}$). When $\arg(\Delta) = \arg(\pm \alpha)$, $\arg(W_{j}) = \arg(\Delta)$ becomes momentum-independent, so the relative phase between components of $\v{u}_{\ell}$ is constant in momentum at linear order. As a result, $\widetilde{\Omega}_{\ell}(\v{q})=0$ at quadratic order, explaining the appearance of Berry curvature zeros approximately along $\arg(\Delta) = \arg(\pm\alpha)$ lines in Fig.~\ref{fig:large_vF_berry_distrib}(b).

\section{Emergent symmetries of the three-band model}
\label{sec:emergent_symmetries}
The three-band Hamiltonian in Eq.~\eqref{eq:ham_3x3_expression} can be tuned to possess several emergent symmetries that become apparent within a rotated basis. We derive the conditions for their emergence and unpack their implications in  Sec.~\ref{sec:lim_topo_flat_bands} and Sec.~\ref{sec:lim_triv_flat_bands}.
 
A change of basis exposes the symmetry structure of  $H$. We apply the following unitary:
\begin{equation}
\label{eq:P_omega}
    P_\omega = \frac{1}{\sqrt{3}} 
    \begin{pmatrix}
        1 & 1 & 1 \\
        1 & \omega & \overline{\omega} \\
        1 & \overline{\omega} & \omega
    \end{pmatrix}.
\end{equation}
At $\v{q} = 0$, this basis coincides with $\ket{\ell = 0}$, $\ket{\ell = -1}$, and $\ket{\ell = 1}$ $C_3$ eigenstates. The transformed Hamiltonian is given by
\begin{equation}
    \label{eq:H_omega}
    H_\omega = P_\omega H P_\omega^\dagger = 
    \begin{pmatrix}
        \Delta_1 & \overline{J}_{5}(q) & J_{3}(q) \\
        J_{5}(q) & \Delta_3 & \overline{J}_{1}(q) \\
        \overline{J}_{3}(q) & J_{1}(q) & \Delta_5
    \end{pmatrix},
\end{equation}
where the diagonal terms are
\begin{equation}
\label{eq:delt_j_def}
    \Delta_j = 2 \text{Re}(\omega^{j - 1} \Delta),
\end{equation}
and the off-diagonal couplings are given by
\begin{equation}
\label{eq:mom_couplings}
    J_{m}(q) = \frac{q}{2} (v_F + \omega^{m-1} \alpha + \overline{\omega}^{m-1} \overline{\alpha}).
\end{equation}
In the above, $\overline{J}_{m}(q) = J_{m}(\overline{q})$. The form of Eq.~\eqref{eq:H_omega} makes clear the roles of the system's parameters: $\Delta$ serves as an effective mass term at the $C_3$-invariant origin, and $v_F, \alpha$ are the momentum-dependent couplings that hybridize bands and generate Berry curvature.

In $H_\omega$, when $J_m = 0$ for some $m$, the Hamiltonian obtains a $U(1)$-symmetry. Consider the case of $J_3 = 0$:
\begin{equation}
    \label{eq:U(1)_sample_ham}
    H_\omega = \begin{pmatrix}
        \Delta_1 & \overline{J}_5(q) & 0 \\
        J_5(q) & \Delta_3 & \overline{J}_1(q) \\
        0 & J_1(q) & \Delta_5
    \end{pmatrix}.
\end{equation}
This has the following $U(1)$ symmetry generated by $Q_3 = \text{diag}(-1, 0, 1)$:
\begin{equation}
\label{eq:U(1)_def}
    e^{-i \psi Q_m} H_\omega(R_\psi \v{q}) e^{i \psi Q_m} = H_\omega(\v{q}),
\end{equation}
where $R_\psi$ acts as a counterclockwise rotation of $\v{q}$ by angle $\psi$ and we write $Q_m$ in place of $Q_3$. Setting $J_5$ or $J_1$ to zero similarly gives rise to $U(1)$ symmetries generated by $Q_5 = \text{diag}(1, -1, 0)$ and $Q_1 = \text{diag}(0, 1, -1)$ respectively. 

Observe that when 2 of $J_1, J_3, J_5$ simultaneously vanish, $H_\omega$ possesses a $U(1) \times U(1)$ symmetry generated by 2 of $Q_1, Q_3, Q_5$. The component of this $U(1) \times U(1)$ that is diagonal in $q$ acts as
\begin{equation}
    e^{-i \psi (Q_m - Q_{m'})} H_\omega(\v{q}) e^{i \psi (Q_m-Q_{m'})}= H_\omega(\v{q}).
\end{equation}

The Hamiltonian in addition possesses an anti-unitary symmetry $P K$ if it satisfies $\Delta_j = \Delta_k, J_j = \pm J_k$ for some $j \neq k$. This anti-unitary symmetry acts on the Hamiltonian as follows:
\begin{equation}
\label{eq:PK_def}
    (P K) H_\omega (K P) = P \overline{H}_\omega P = H_\omega.
\end{equation}
 In the above, $P$ is a real matrix that depends on the choice of $j, k$ as well as the relative sign between $J_j, J_k$. For example, when $j=1, k = 5$, $H_\omega$ and $P$ are given by
\begin{equation}
\label{eq:PK_sample_ham}
    H_\omega^\pm = \begin{pmatrix}
        \widetilde{\Delta} & J \overline{q} & J_3(q) \\
        J q & \Delta_3 & \pm J \overline{q} \\
        \overline{J}_3(q) & \pm J q & \widetilde{\Delta}
    \end{pmatrix}, \quad
    P_\pm = \begin{pmatrix}
        0 & 0 & \pm 1 \\
        0 & 1 & 0 \\
        \pm 1 & 0 & 0
    \end{pmatrix}, 
\end{equation}
where $\widetilde{\Delta} \equiv \Delta_1 = \Delta_5$ and the superscript labels the relative sign between $J_1$ and $J_5$. This anti-unitary symmetry enforces vanishing spinorless Berry curvature over all bands of $H_\omega$.

\begin{table}[h]
  \centering
  \begin{tabular}{|c|c|c|}
    \hline
    {Symmetry} & {Conditions on $J_m, \Delta_j$} & {Conditions on $\Delta, \alpha, v_F$} \\ \hline
    $U(1)$ &  $J_m = 0$ &  $v_F = -2 \text{Re}(\omega^{m-1} \alpha)$\\ \hline 
    \multirow{2}{*}{$P K$}& $ \Delta_j = \Delta_k$ & $\Delta = \pm |\Delta| e^{i \frac{\pi}{3} (2 - j - k)}$ \\ \cline{2-3}
     & $ J_j = J_k$ & $0 = \mathrm{Re}(\alpha [\omega^{j-1} - \omega^{k-1}])$ \\ 
     & \uline{or} & \uline{or} \\
     & $J_j = - J_k$ & $v_F = \mathrm{Re}(\alpha \omega^{2 - j - k})$
     \\ \hline
  \end{tabular}
  \caption{Necessary criteria for the emergence of $U(1)$ and $PK$ symmetries of $H_\omega$. The condition needs to be satisfied for some $m$ or $j\neq k$.}
  \label{tab:emergence_crit}
\end{table}

Table~\ref{tab:emergence_crit} summarizes the conditions on matrix elements $\Delta_n$, $J_m$ of $H_\omega$ for the symmetries $U(1), PK$ to emerge.

Furthermore, the criterion for the emergence of a $U(1) \times U(1)$ symmetry can be understood as a condition on $\alpha$. For instance, $J_1 = J_3 = 0$ implies
\begin{equation}
    \begin{split}
        -2 \text{Re}(\alpha) &= -2 \text{Re}(\overline{\omega} \alpha) \implies \arg(\alpha) \in \left \{- \frac{2 \pi}{3}, \frac{\pi}{3} \right \}.
    \end{split}
\end{equation}
Analogous conditions can be derived when $J_1 = J_5 = 0$ or $J_3 = J_5 =0$.

These conditions also explain a feature left open in Sec.~\ref{sec:spinorless_phenomenology}:
the change in the structure of Fig.~\ref{fig:spinorless_ground_bc_signs} at
$v_F/|\alpha| = 2$. Beyond that ratio, $J_m \neq 0$ for every $\arg(\alpha)$, so
no $U(1)$ symmetry can emerge.

\section{Emergent symmetry with a topological flat band}
\label{sec:lim_topo_flat_bands}

\begin{figure*}
    \centering
    \includegraphics[width=\linewidth]{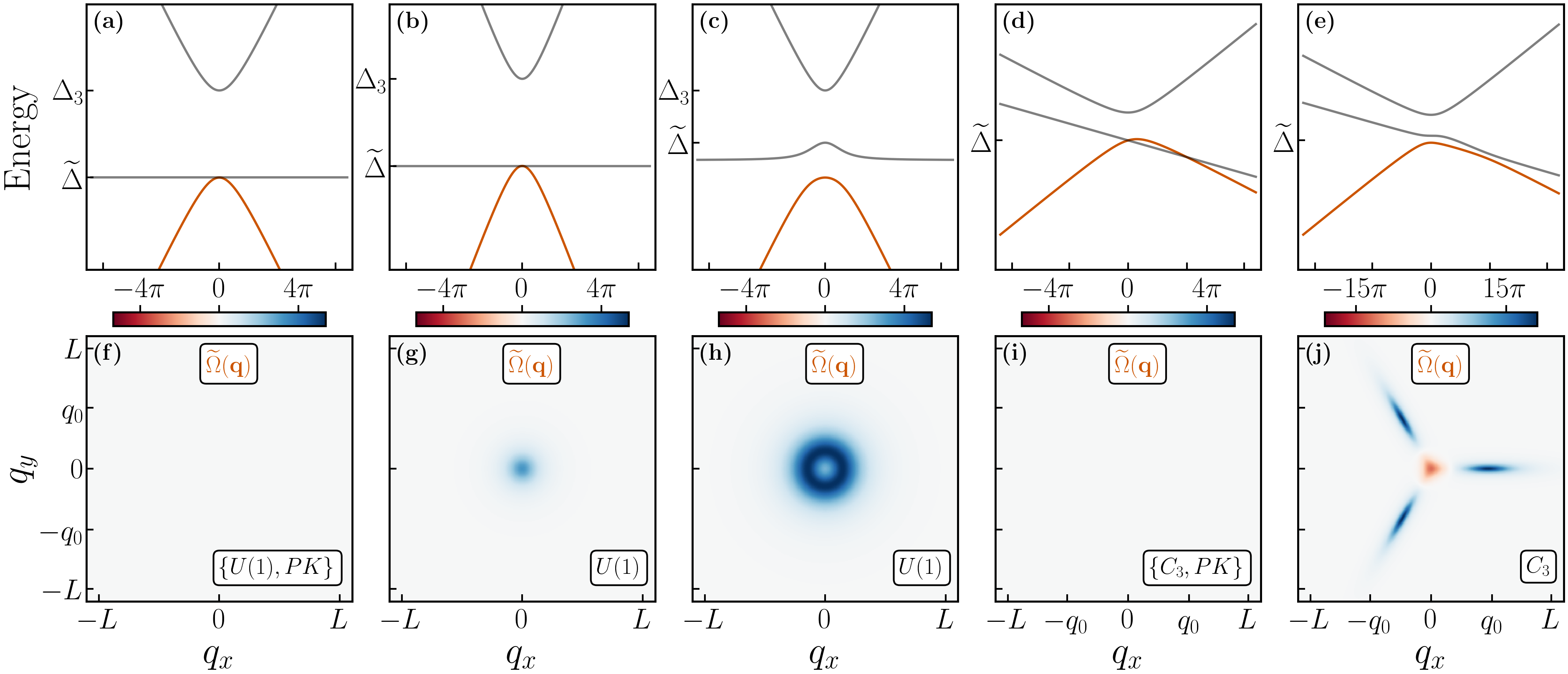}
    \caption{
    Band structures and associated spinorless Berry curvature of ground state of $H_\omega$ in several different regimes. For all plots, $L = 1.1$. (a-e) show band structure when the symmetries are $\{U(1), PK\}$, $\{U(1)\}$, $\{U(1)\}$, $\{C_3, PK \}$, $\{C_3\}$ respectively. (f-j) show spinorless Berry curvature when the symmetries are $\{U(1), PK\}$, $\{U(1)\}$, $\{U(1)\}$, $\{C_3, PK \}$, $\{C_3\}$ respectively. (a, f) use $\{\Delta = \omega/4, \alpha=\omega, v_F = -2\text{Re}(\overline{\omega} \alpha)\}$. (b, g) use $\{\Delta = \omega/4, \alpha = \omega-1/2, v_F = -2\text{Re}(\overline{\omega} \alpha)\}$. (c, h) use $\{\Delta = \omega/4 + 1/10, \alpha = \omega, v_F = -2\text{Re}(\overline{\omega} \alpha) \}$. (d, i) use $\{\Delta = 0.08 \omega, \alpha = \omega/2, v_F = -2 \text{Re}(\overline{\omega} \alpha) + 0.57 \}$. (e, j) use $\{\Delta = 0.08 \omega + 0.02, \alpha = \omega/2, v_F = -2 \text{Re}(\overline{\omega} \alpha) + 0.57 \}$.}
    \label{fig:topo_flat_band_phenom}
\end{figure*}

In this section, we show that when $U(1)$ and $PK$ symmetries coexist, the 3-band Hamiltonian can host a \textit{singular} topological flat band. This topological flat band contains an irremovable singularity at $\v{q} = 0$, so the band can only obtain a Chern number if its exact flatness is broken \cite{SFB_wang_yao, sfb_classification, sfb_frustrated_hopping}. We derive the condition for its appearance, and analyze the system's behavior upon adding perturbations (see Fig.~\ref{fig:topo_flat_band_phenom}).

We now consider the situation in which the Hamiltonian possesses both an emergent $U(1)$-symmetry and an anti-unitary $PK$ symmetry.

\subsection{Quadratic band touching with \texorpdfstring{$U(1)$}{U(1)} and anti-unitary symmetries}
\label{sec:QBT_U(1)_PK}

We begin at a point where the Hamiltonian is $U(1)$-symmetric and has the anti-unitary symmetry $PK$. By tuning $\arg(\Delta) = 2 \pi/3$, $v_F = -2 \text{Re}(\overline{\omega} \alpha)$, and $\arg(\alpha) = 2 \pi/3$,
we obtain the Hamiltonian
\begin{equation}
\label{eq:U(1)_PK_H}
   H_\omega = \begin{pmatrix}
        \widetilde{\Delta} & J \overline{q} & 0 \\
        J q & \Delta_3 & J \overline{q} \\
        0 & J q & \widetilde{\Delta}
    \end{pmatrix}.
\end{equation}
The $U(1)$-symmetry generator is given by $Q_3 = \text{diag}(-1, 0, 1)$. Additionally, since $\Delta_1 = \Delta_5$ and $J_1 = J_5$, the Hamiltonian is symmetric under the anti-unitary symmetry generated by $P_+$ in Eq.~\eqref{eq:PK_sample_ham}.
The anti-unitary symmetry enforces vanishing Berry curvature across all of k-space, which is visible in Fig.~\ref{fig:topo_flat_band_phenom}(f).

This Hamiltonian possesses a singular topological flat band at energy $\widetilde{\Delta}$ with a quadratic band touching (Fig.~\ref{fig:topo_flat_band_phenom}(a)).
The energy and the spinor of the topological flat band are
\begin{equation}
\label{eq:U(1)_PK_ev0}
    \epsilon_0 = \widetilde{\Delta}, \v{u}_0 = (\overline{q}, 0, -q)^T.
\end{equation}
At $\v{q}=0$, this flat band is degenerate with the lower band.
Furthermore, $\v{u}_0$ has twice the phase winding of a Dirac-like spinor. Thus the gapless point at $q = 0$ is a quadratic band touching (QBT)\footnote{This quadratic band touching between a topological flat band and a dispersive band is also found in the singular flat band model proposed in Ref.~\cite{ SFB_wang_yao}.}.

\subsection{Breaking anti-unitary orbital exchange}
\label{sec:QBT_U(1)}

We now consider perturbing Eq.~\eqref{eq:U(1)_PK_H} to break the $PK$ symmetry. Let us first consider setting $J_1 \neq J_5$ while keeping $J_3 = 0$, which can be accomplished by perturbing $\arg(\alpha)$ away from $2 \pi/3$. The resulting Hamiltonian is given by
\begin{equation}
\label{eq:U(1)_H}
    H_\omega = \begin{pmatrix}
        \widetilde{\Delta} & M \overline{q} & 0 \\
        M q & \Delta_3 & P \overline{q} \\
        0 & P q & \widetilde{\Delta}
    \end{pmatrix}
\end{equation}
for $M, P \in \mathbb{R}$. The Hamiltonian still possesses a topological flat band at
\begin{equation}
\label{eq:U(1)_ev0}
    \epsilon_0 = \widetilde{\Delta}, \v{u}_0 = (-P \overline{q}, 0, M q)^T,
\end{equation}
which can be seen in Fig.~\ref{fig:topo_flat_band_phenom}(b).
Despite the lack of $PK$ symmetry, $\v{u}_0$ possesses no spinorless Berry curvature since the relative weight $- \frac{P \overline{q}}{M q}$ does not depend on $|\v{q}|$.
The dispersive partners are not protected, however: the same PK-breaking
imbues the top and bottom bands with finite curvature as shown in Fig.~\ref{fig:topo_flat_band_phenom}(g).

\subsection{Gapped QBT as a source of Berry curvature}
\label{sec:gapped_QBT}
We now break the $PK$ symmetry by tuning $\arg(\Delta)$ away from $2\pi/3$.
In the transformed basis for $H_\omega$, this breaks the degeneracy between $\Delta_1$ and $\Delta_5$. The resulting spinorless Berry curvature in Fig.~\ref{fig:topo_flat_band_phenom}(h) is annular: a signature of a gapped quadratic band touching. 

\subsection{Splitting QBT into vorticity-conserving Dirac cones}
\label{sec:topo_dirac_cones}
Breaking the $U(1)$-symmetry down to $C_3$ results in the quadratic band touching splitting into a set of 4 symmetry-protected massless Dirac cones. By adding $M q, M \overline{q}$ components to $H_\omega$ in Eq.~\eqref{eq:U(1)_PK_H}, we arrive at the Hamiltonian
\begin{equation}
\label{eq:gapless_dirac_cones_H}
    H_\omega = \begin{pmatrix}
        \widetilde{\Delta} & J \overline{q} & M q \\
        J q & \Delta_3 & J \overline{q} \\
        M \overline{q} & J q & \widetilde{\Delta}
    \end{pmatrix}.
\end{equation}
The Hamiltonian preserves the $PK$ symmetry.
Along the $q_x$-axis, we can diagonalize $H_\omega$ to obtain (see App.~\ref{app:C3_PK})
\begin{gather}
\label{eq:C3_PK_ev0}
    \epsilon_0 = \widetilde{\Delta} - M q_x, 
    \\
    \epsilon_{\pm} = \frac{(\Delta_3 + \widetilde{\Delta} + M q_x)}{2} \pm \frac{1}{2} \sqrt{(\Delta_3 - \widetilde{\Delta} - M q_x)^2 + 8 J^2 q_x^2 }.
\end{gather}
The spectrum has two gapless points at $(q_x, q_y) = (0, 0)$ and $(q_x, q_y) = (\frac{M [\Delta_3 - \widetilde{\Delta}]}{J^2 - M^2}, 0)$, which can be rotated by $\pm 2 \pi /3$ to identify all 4 gapless points. Breaking $U(1) \rightarrow C_3$ splits the QBT into 4 massless Dirac cones such that the 3 satellite cones wind opposite to the cone at $(q_x, q_y) = (0, 0)$, conserving the total winding.

\subsection{Massive Dirac cones as a source of Berry curvature}

Breaking the $P K$ symmetry of Eq.~\eqref{eq:gapless_dirac_cones_H} generates nonzero Berry curvature. Consider perturbing the Hamiltonian with $\delta V$ given by
\begin{equation}
\label{eq:gapped_dirac_cones_pert}
    \delta V = \begin{pmatrix}
        \widetilde{\Delta}_1 & 0 & 0 \\
        0 & \widetilde{\Delta}_3 & 0 \\
        0 & 0 & \widetilde{\Delta}_5
    \end{pmatrix}.
\end{equation}
At $(q_x, q_y) = (0, 0)$, we project $\delta V$ into the subspace spanned by $\v{u}_-, \v{u}_0$ and determine the effective mass of the Dirac cone to be
\begin{equation}
\label{eq:kappa_dirac_mass}
    m_*(q_x=0) = \frac{\widetilde{\Delta}_1 - \widetilde{\Delta}_5}{2}.
\end{equation}
The perturbation $\delta V$ also gaps out the satellite Dirac cones shown in Fig.~\ref{fig:topo_flat_band_phenom}(d), resulting in nonzero Berry curvature (see Fig.~\ref{fig:topo_flat_band_phenom}(j)). The relative sign of the Berry curvature at the origin and the satellite Dirac cones is not fixed, as is evident from the discussion in Sec.~\ref{sec:spinorless_finitemomentum}.

\section{Emergent symmetry with a trivial flat band}
\label{sec:lim_triv_flat_bands}
\begin{figure*}
    \centering
    \includegraphics[width=\linewidth]{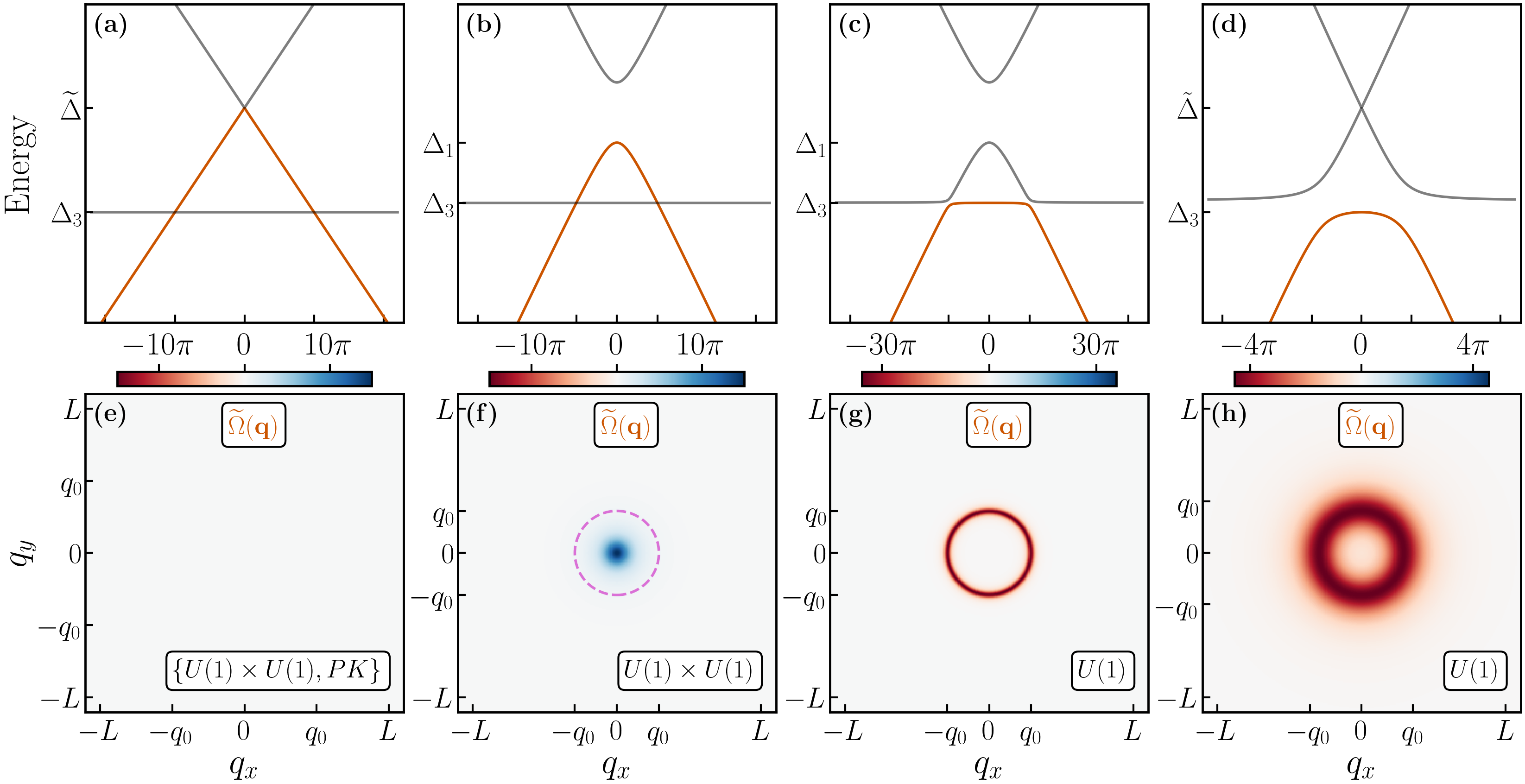}
    \caption{Band structures and associated spinorless Berry curvature of blue band of $H_\omega$ in several different regimes. For all plots, $L = 1.1$. (a-d) show band structure when the symmetries are $\{U(1) \times U(1), PK\}$, $\{U(1) \times U(1)\}$, $\{U(1)\}$, $\{U(1) \}$ respectively. (e-h) show spinorless Berry curvature when the symmetries are $\{U(1) \times U(1), PK\}$, $\{U(1) \times U(1)\}$, $\{U(1)\}$, $\{U(1) \}$ respectively. (a, e) use $\{ \Delta = -\omega/4, \alpha = \omega, v_F = -2 \text{Re}(\omega \alpha) \}$. (b, f) use $\{\Delta = -i/4, \alpha = 1.4 \omega, v_F = -2 \text{Re}(\omega \alpha) \}$. (c, g) use $\{ \Delta = -i/4, \alpha = 1.4 \omega + 0.1, v_F = -2 \text{Re}(\omega \alpha) \}$. (d, h) use $\{ \Delta = -\omega/4, \alpha = 1.4 \omega + 0.5, v_F = -2 \text{Re}(\omega \alpha) \}$.
    }
    \label{fig:triv_flat_band_phenom}
\end{figure*}

In this section, we focus on the consequence of $U(1) \times U(1)$ symmetry as well as an anti-unitary $P K$ symmetry. Concretely, the Hamiltonian we begin with is given in Eq.~\eqref{eq:U(1)_U(1)_PK_H}. 
\begin{equation}
\label{eq:U(1)_U(1)_PK_H}
    H_\omega = \begin{pmatrix}
        \widetilde{\Delta} & 0 & J q \\
        0 & \Delta_3 & 0 \\
        J \overline{q} & 0 & \widetilde{\Delta}
    \end{pmatrix}.
\end{equation}
This Hamiltonian possesses 2 independent $U(1)$ generators $Q_5 = \text{diag}(1, -1, 0)$ and $Q_1 = \text{diag}(0, 1, -1)$. Its anti-unitary symmetry is given by $P_+ K$ since $J_1 = J_5$ and $\Delta_1 = \Delta_5$. The $U(1) \times U(1)$ symmetry decouples the eigenvector $\v{u}_0 = (0, 1, 0)^T$ of $H_\omega$ from other orbitals.
In the following sections, we discuss the band structure and Berry curvature distributions that appear when we perturb $H_\omega$.

\subsection{Massless Dirac cone with \texorpdfstring{$U(1) \times U(1)$}{U(1)xU(1)} and anti-unitary symmetries}

We first analyze the Hamiltonian in Eq.~\eqref{eq:U(1)_U(1)_PK_H}.
Its eigenvalues and un-normalized eigenvectors of $H_\omega$ are
\begin{equation}
\label{eq:highly_sym_crit_eigen}
    \begin{split}
    \epsilon_0 &= \Delta_3, \v{u}_0 = (0, 1, 0)^T \\
        \epsilon_- &= \widetilde{\Delta} - |J| |q|, \v{u}_- = (|J| |q|, 0, -J \overline{q})^T \\
        \epsilon_+ &= \widetilde{\Delta} + |J| |q| , \v{u}_+ = (|J| |q|, 0, J \overline{q})^T.
    \end{split}
\end{equation}
The eigenvectors $\v{u}_\pm$ together form a massless Dirac cone at $\v{q} = 0$.
The eigenvector $\v{u}_0$ is a topologically trivial flat-band (Fig.~\ref{fig:triv_flat_band_phenom}(a)). $\epsilon_\pm$ and $\epsilon_0$ intersect at \begin{equation}
\label{eq:U(1)_U(1)_PK_q0}
    |q_0| = \frac{| \Delta_3 - \widetilde{\Delta}|}{|J|},
\end{equation}
giving rise to a ring of degeneracies.

\subsection{Breaking anti-unitary orbital exchange}

Next, we break $P K$ symmetry by perturbing Eq.~\eqref{eq:U(1)_U(1)_PK_H} with $\text{diag}(\Delta_1', \Delta_3', \Delta_5')$, which yields the Hamiltonian
\begin{equation}
\label{eq:U(1)_U(1)_H}
H_\omega = \begin{pmatrix}
    \Delta_1 & 0 & J q \\
    0 & \Delta_3 & 0 \\
    J \overline{q} & 0 & \Delta_5
\end{pmatrix}.
\end{equation}
It retains a topologically trivial flat band
\begin{equation}
\label{eq:U(1)_U(1)_ev0}
    \epsilon_0 = \Delta_3, \v{u}_0 = (0, 1, 0)^T.
\end{equation}
The remaining two bands together span a Dirac cone with an effective mass
\begin{equation}
    m_* = \frac{\Delta_1 - \Delta_5}{2}.
\end{equation}
The Berry curvature of $\v{u}_-$ is shown in Fig.~\ref{fig:triv_flat_band_phenom}(f), where the pink circle indicates the ring of degeneracy between $\epsilon_0$ and $\epsilon_-$. The radius of this ring is
\begin{equation}
\label{eq:avoided_radius}
    |q_0| = \frac{\sqrt{|\Delta_3^2 + \Delta_1 \Delta_5 - \Delta_3 \Delta_1 - \Delta_3 \Delta_5|}}{|J|}.
\end{equation}
Observe that when either $\Delta_1 = \Delta_3$ or $\Delta_3 = \Delta_5$, the radius $|q_0| = 0$, so the gapless ring shrinks to a point.

\subsection{Splitting \texorpdfstring{$U(1) \times U(1) \rightarrow U(1)$}{U(1)xU(1)}}
\label{sec:splitting_U(1)_U(1)}

We next consider breaking $U(1) \times U(1) \rightarrow U(1)$ by introducing off-diagonal elements $Mq, M \overline{q}$ to the Hamiltonian in Eq.~\eqref{eq:U(1)_U(1)_H} via the perturbation
\begin{equation}
\label{eq:U(1)_U(1)_split_pert}
    \delta V = \begin{pmatrix}
        0 & 0 & 0 \\
        0 & 0 & M \overline{q} \\
        0 & M q & 0 
    \end{pmatrix}.
\end{equation}
The term $\delta V$ can be added to the Hamiltonian by perturbing $\alpha \rightarrow \alpha + 2M/3$ and $v_F \rightarrow v_F + 2M/3$ for $M \in \mathbb{R}$. This perturbation breaks the ring of degeneracy between $\epsilon_0$ and $\epsilon_-$, producing annular Berry curvature shown in Fig.~\ref{fig:triv_flat_band_phenom}(g).

We can in fact tune $\Delta$ to make the upper two bands gapless. The Berry curvature ring still persists even in that scenario, see Fig.~\ref{fig:triv_flat_band_phenom}(h).

\section{Un-truncated model}
\label{sec:connection_to_full_model}

Thus far, we have focused on the band structure of $H$ and on the spinorless contribution to the total Berry curvature of the three-band Hamiltonian. We now investigate the behavior of the full model and identify when the spinorless Berry curvature captures its phenomenology.

We first identify regimes where the spinorless Berry curvature provides a good approximation to the total Berry curvature of the three-band model. We then determine the regions of momentum space and parameter ranges in which the three-band model accurately reproduces the energy and Berry curvature of the low-lying bands of an un-truncated model.
Taken together, this analysis identifies the regimes in which the three-band Hamiltonian's spinorless Berry curvature is reliable. In these regimes it reproduces, near the $\v{\kappa}$ point, the momentum dependence of the total Berry curvature of the untruncated model's low-energy states.

\subsection{Parent spinors}
\label{sec:par_spinors}
We consider the 2-component $\lambda$-jellium parent spinors proposed in \cite{lambda_jellium}:
\begin{equation}
    | \chi(\v{q}) \rangle = \frac{1}{\sqrt{1 + \lambda^2 (q_x^2 + q_y^2)}}\begin{pmatrix}
        1 \\
        \lambda (q_x + i q_y)
    \end{pmatrix}.
    \label{eq:lambda_jell}
\end{equation}
Here $\lambda > 0$ tunes the concentration of Berry curvature around $\v{q} = 0$. The spinor is holomorphic in $q = q_x + i q_y$ and remains invariant under the transformation $q \rightarrow \overline{q}, \chi^a \rightarrow \overline{\chi}^a$.
The Berry curvature of the $\lambda$-jellium spinor is
\begin{equation}
    \Omega_\text{orig}(\v{k}) = -2 \bigg( \frac{\lambda}{1 + \lambda^2 (k_x^2 + k_y^2)} \bigg)^2
\end{equation}
such that the parent contribution (Eq.~\eqref{eq:parent_contrib_def}) at $\v{q} = 0$ is
\begin{equation}
\begin{split}
    \Omega^\text{par}_n(\v{q} = 0) &= \sum_{j \in \{1, 3, 5\}} |u_n^j(\v{q}=0)|^2 \Omega_\text{orig}(\v{\kappa}_j) \\
    &= -2 \bigg(\frac{\lambda}{1 + \lambda^2 |\v{\kappa}|^2} \bigg)^2
\end{split}
\end{equation}
We focus on a class of periodic potentials whose orbital-space action is $\hat{V} = \text{diag}(v_1, v_2)$ for $v_1, v_2 \in \mathbb{R}$. Under this potential, we have
\begin{equation}
\label{eq:delta_alpha}
    \begin{split}
        \Delta &= \frac{v_1 + v_2 \omega \lambda^2 |\v{\kappa}|^2}{2(1 + \lambda^2 |\v{\kappa}|^2)}  \\
\alpha &= \frac{|\v{\kappa}| \lambda^2}{4 (1 + \lambda^2 |\v{\kappa}|^2)^2} \bigg( 2 v_1 + v_2[4 \overline{\omega} - 2 \sqrt{3} e^{i \pi /6} |\v{\kappa}|^2 \lambda^2] \bigg).
    \end{split}
\end{equation}

The parameterization specified by Eq.~\eqref{eq:delta_alpha} provides a means of accessing the emergent features discussed in Sec.~\ref{sec:lim_topo_flat_bands} and Sec.~\ref{sec:lim_triv_flat_bands} by tuning the physical parameter $\lambda$. For $v_1 = v_2 = -1$, the phase of $\Delta$ crosses $-2\pi/3$ precisely at $\lambda_\mathrm{crit} = 1/|\v{\kappa}|$, where the two lowest bands at $\v{\kappa}$, carrying $C_3$ angular momenta $\ell = 0$ and $\ell = -1$, become degenerate and the system undergoes band inversion. We therefore refer to $\lambda < \lambda_\mathrm{crit}$ as \emph{subcritical}, $\lambda > \lambda_\mathrm{crit}$ as \emph{supercritical}, and $\lambda \approx \lambda_\mathrm{crit}$ as the \emph{critical regime}.

\subsection{Scaling of Berry curvature near band inversion}
\label{sec:bc_scaling}

\begin{figure}
    \centering
    \includegraphics[width=\linewidth]{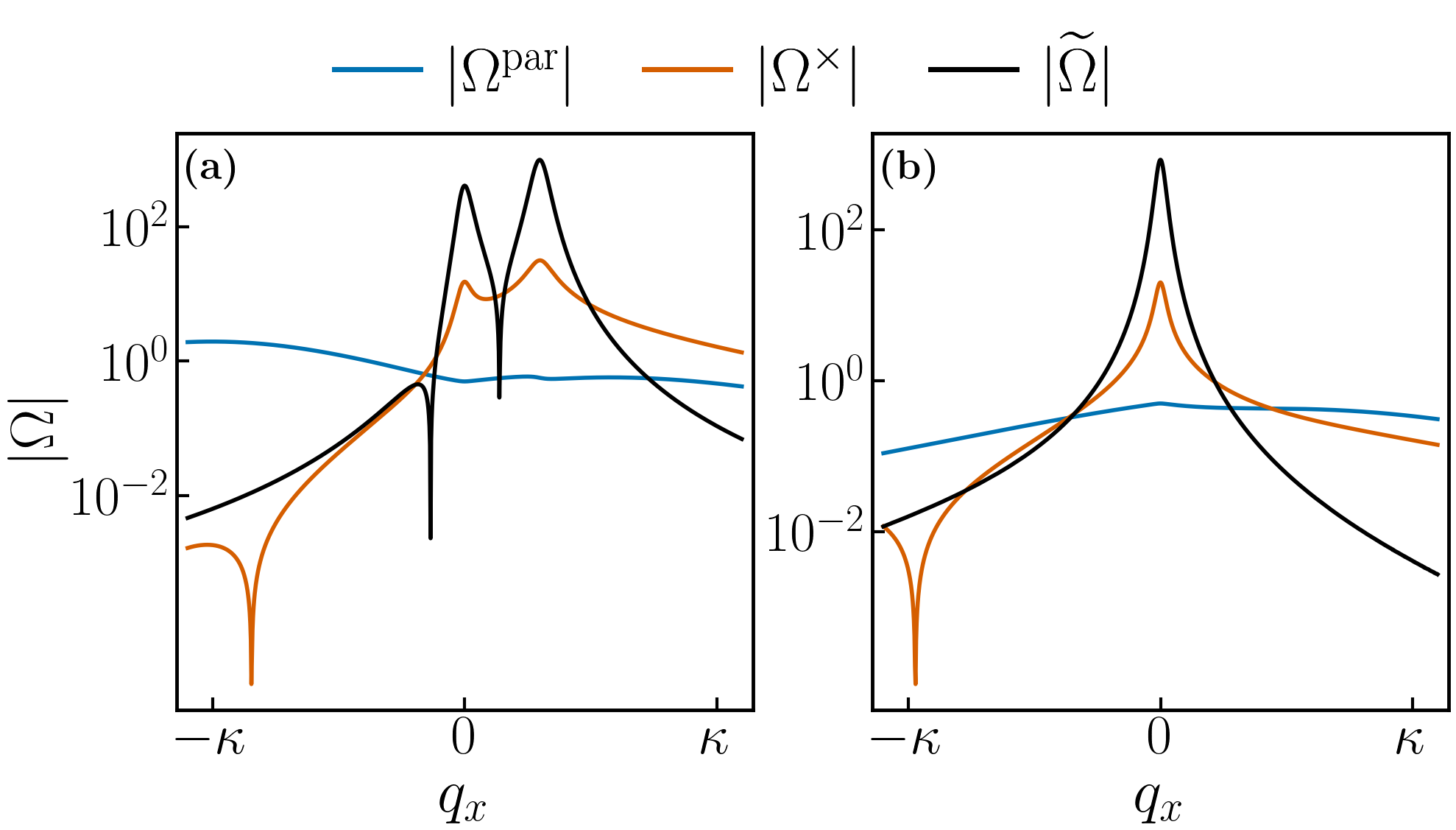}
    \caption{Scaling of Berry curvature contributions. The potential is $v_1 = v_2 = -1$ and $\lambda = 0.99/|\v{\kappa}|$. (a) shows scalings when $v_F/|\alpha| = 10^{-0.5} < 2$. (b) shows scalings when $v_F/|\alpha| = 10^{0.5} > 2$.
    }
    \label{fig:near_crit_bc_scaling}
\end{figure}

In this section, we show that $\widetilde{\Omega}$ captures the behavior of $\Omega^\text{tot}$ near band inversion. In Fig.~\ref{fig:near_crit_bc_scaling}, we show the three contributions to Berry curvature near criticality (we fix $|\v{\kappa}| = 1$ throughout).  As $q_x$ approaches zero, the spinorless contribution dominates over the other contributions, indicating the relevance of the spinorless approximation. Moreover, the additional peak in Fig.~\ref{fig:near_crit_bc_scaling}(a) can be attributed to the existence of a satellite Dirac cone, which is discussed in Sec.~\ref{sec:topo_dirac_cones}. The extreme dips in the plots in Fig.~\ref{fig:near_crit_bc_scaling}(a, b) are a result of sign changes in Berry curvature. 

We can analytically verify this behavior. Recall that for the potential $v_1 = v_2 = -1$, the 2 $C_3$ eigenvalues of interest are $\ell = 0, -1$. Then we can compare the spinorless and cross term contributions in these respective $\ell$ sectors for $\lambda \approx \lambda_\mathrm{crit}$. The spinorless terms are 
\begin{equation}
    \begin{split}
        \widetilde{\Omega}_{\ell = 0}(0, 0) &\sim \frac{-\lambda^6}{(\lambda^4 - 1)^2} - \frac{\lambda^{12} v_F^2}{(\lambda^4 - 1)^2} \\
            \widetilde{\Omega}_{\ell = -1} (0, 0) &\sim \frac{\lambda^6}{(\lambda^4 - 1)^2} + \frac{\lambda^6 v_F^2}{(\lambda^4 - 1)^2},
    \end{split}
\end{equation}
whereas the cross-terms are given by
\begin{equation}
    \begin{split}
        \Omega^\times_{\ell=0}(0, 0) &\sim -\frac{\lambda^6}{(1 + \lambda^2)^2 |\lambda^2 - 1|} + \frac{\lambda^8 v_F}{(1 + \lambda^2)^2 |\lambda^2 - 1|}\\
        \Omega^\times_{\ell = -1}(0, 0) &\sim \frac{\lambda^6}{(1 + \lambda^2)^2 |\lambda^2 - 1|} - \frac{\lambda^6 v_F}{(1 + \lambda^2)^2 |\lambda^2 - 1|}.
    \end{split}
\end{equation}
Since $\widetilde{\Omega}(0, 0) \sim \frac{1}{(1 - \lambda^4)^2}$ and $\Omega^\times(0, 0) \sim \frac{1}{(1 - \lambda^2)}$, the spinorless Berry curvature dominates for near-critical $\lambda$ ($\lambda \approx 1$). The parent contribution is non-singular at criticality, so it is irrelevant in such regimes. As a result, any near-critical momentum-dependent behavior of the total Berry curvature can be described by spinorless Berry curvature.

\subsection{Quadratic dispersion model}
\label{sec:quadratic_disp}

We now consider the periodic potential
\begin{equation}
\label{eq:un_trunc_moire_pot}
    \hat{V} = \sum_{\v{k}, a, \v{G}, j}  V^a_j(\v{k} + \v{G}) c^\dagger_{\v{k} + \v{G} + \v{g}_j} c_{\v{k} + \v{G}} + \mathrm{h.c.}
\end{equation}
where $V^a_j(\v{k} + \v{G})$ is given in Eq.~\eqref{eq:V_non_uniform}, and $\v{G} = n_1 \v{g}_1 + n_2 \v{g}_2 + n_3 \v{g}_3$ for $n_i \in \mathbb{Z}$. For the numerical studies in this section and the following section, we take a constant orbital-space potential $V^a = -1$, and we assume $|\v{G}| \leq 5 |\v{g}_1|$.

\begin{figure}
    \centering
    \includegraphics[width=\linewidth]{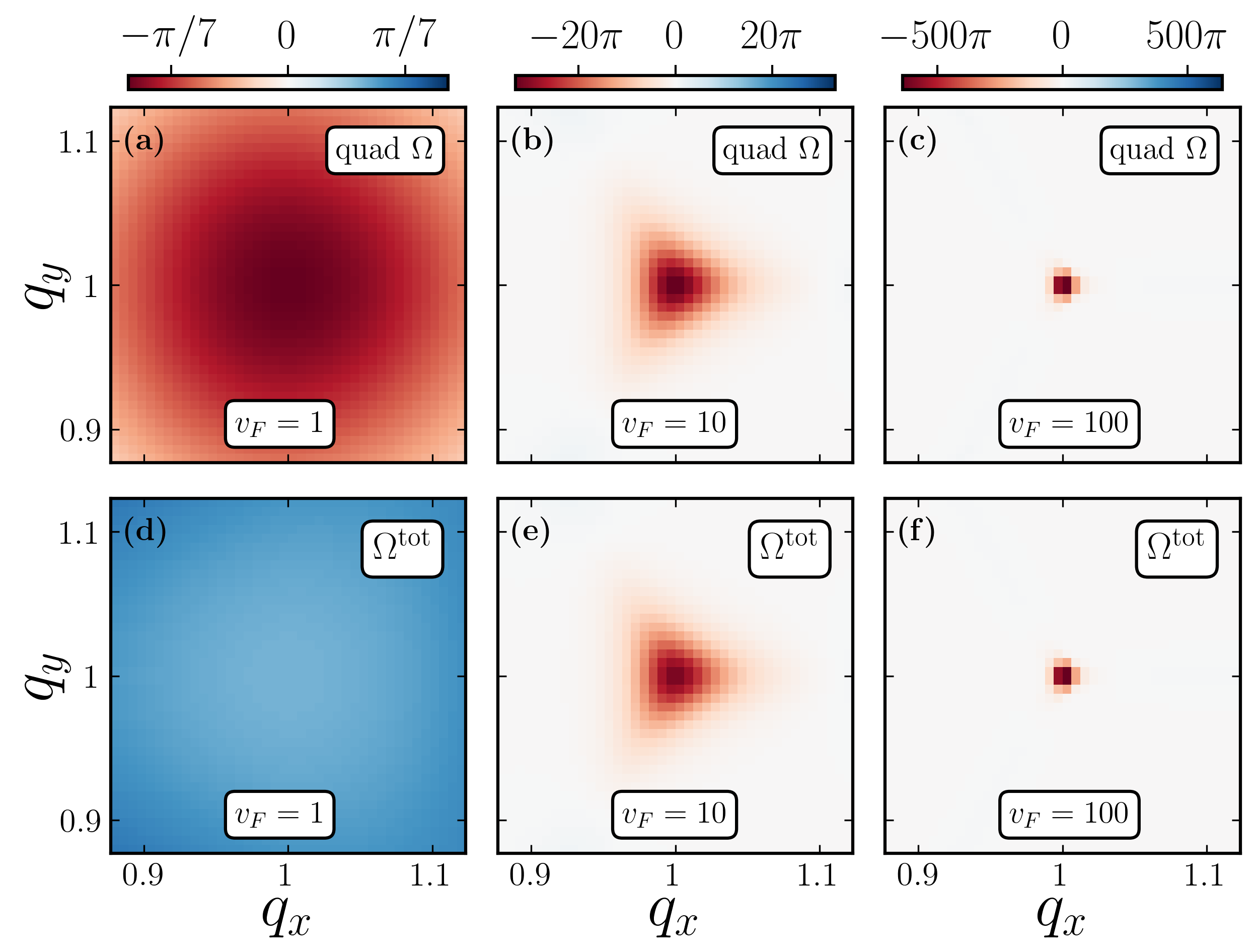}
    \caption{Comparison of Berry curvature of quadratically dispersing model (top row) and total 3-band Berry curvature (bottom). Parameters used are $v_1 = v_2 = -1, |\v{\kappa}| = 1, \lambda = 0.8 / |\v{\kappa}|$. (a-c) show quadratic dispersion Berry curvatures for $v_F = 1, 10, 100$ respectively. (d-f) show total 3-band Berry curvatures for $v_F = 1, 10, 100$ respectively.
    }
    \label{fig:mBZ_benchmarks_0p8}
\end{figure}

We first consider a model with quadratic dispersion
\begin{equation}
\label{eq:quad_disp}
    E(\v{k}) = \frac{v_F}{2 \kappa}|\v{k}|^2 .
\end{equation}
The rotationally-invariant dispersion and the $C_3$ symmetric potential ensure that the spectral weight of low-lying bands will be concentrated around $\v{\kappa}, \v{\kappa}'$. Moreover, large $v_F$ energetically penalizes participation of higher-momentum (beyond 1\textsuperscript{st} BZ) plane-waves in low energy states. Consequently, low-lying eigenstates consist of a superposition of plane-waves at momenta near $\v{\kappa}, \v{\kappa}'$. This means that the 3-band model can reproduce the low-energy physics of the un-truncated model when $v_F \gg |\alpha|$ in both the subcritical ($\lambda < 1/|\v{\kappa}|$) regime (see Fig.~\ref{fig:mBZ_benchmarks_0p8}) and the supercritical ($\lambda > 1/|\v{\kappa}|$) regime (see Fig.~\ref{fig:mBZ_benchmarks_1p2}). Moreover, we find that the spinorless Berry curvature alone accurately captures the true Berry curvature of the quadratically dispersing model ($\mathrm{quad} \ \Omega$) when $v_F \gg |\alpha|$ (see Fig.~\ref{fig:mBZ_benchmarks_component_line_cuts}).

\par
\begin{figure}
    \centering
    \includegraphics[width=\linewidth]{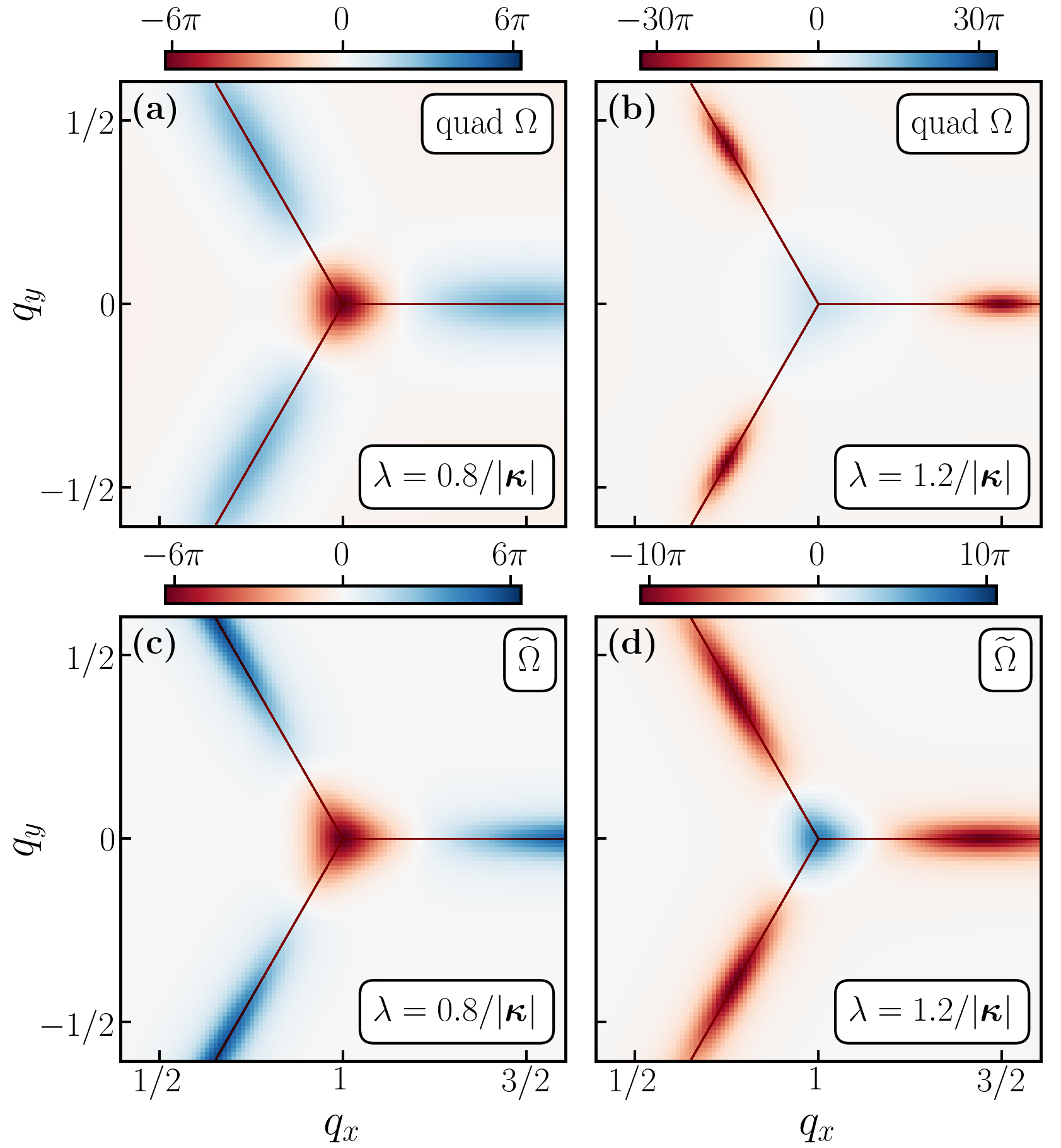}
    \caption{
    Berry curvature for the quadratically dispersing model and the spinorless 3-band model. Here maroon lines indicate BZ boundaries, $v_F/ |\alpha| \approx 7$, and $|\v{\kappa}| = 1$. (a,b) shows quadratic dispersion Berry curvature for $\lambda = 0.8/|\v{\kappa}|, 1.2/|\v{\kappa}|$ respectively. (c,d) shows spinorless Berry curvature for $\lambda = 0.8/|\v{\kappa}|, 1.2/|\v{\kappa}|$ respectively.}
    \label{fig:n=2_bc}
\end{figure}

 Even when $v_F \gg |\alpha|$ is not satisfied, the spinorless Berry curvature from the three-band model can capture qualitative features of the quadratically dispersing model. For $v_F / |\alpha| \approx 7$, the Berry curvature around $\v{\kappa}$ takes the form of a hotspot with one sign surrounded by three $C_3$-related legs of Berry curvature with different signs (Fig.~\ref{fig:n=2_bc}(a, b)). This can be reproduced by the spinorless Berry curvature (Fig.~\ref{fig:n=2_bc}(c, d)).

\subsection{n-dispersion model}
\label{sec:near_crit_behavior}
The emergent $U(1)$ symmetry only appears when $v_F / |\alpha| \leq 2$. The quadratic dispersion cannot attain this ratio while keeping the wavefunction weight concentrated to $\v{\kappa}_{1, 3, 5}$.  To access this regime, we consider the following dispersion~\cite{RMG_dispersion}, which we call the \textit{n-dispersion} model:
\begin{equation}
\label{eq:n_disp}
    E(\v{k}, v_F, m, n) = \bigg(\frac{\sqrt{|\v{\kappa}|^{2n} + m^2}}{n |\v{\kappa}|^{2n-1}} \bigg) v_F \sqrt{|\v{k}|^{2n} + m^2}.
\end{equation}
Here $m$ is a positive real number and $n$ is a positive integer. The prefactor ensures that the linearized dispersion in the vicinity of the $\v{\kappa}$ point is $v_F |\v{k}|$.
At large $n$, the dispersion has a flat band bottom with highly dispersive band at large momentum, penalizing mixing with higher momentum states. We find that $n = 11, m = 10 |\v{\kappa}|^n$ is sufficient to recover the desired $U(1)$ symmetric behavior. 

\begin{figure}
    \centering
    \includegraphics[width=\linewidth]{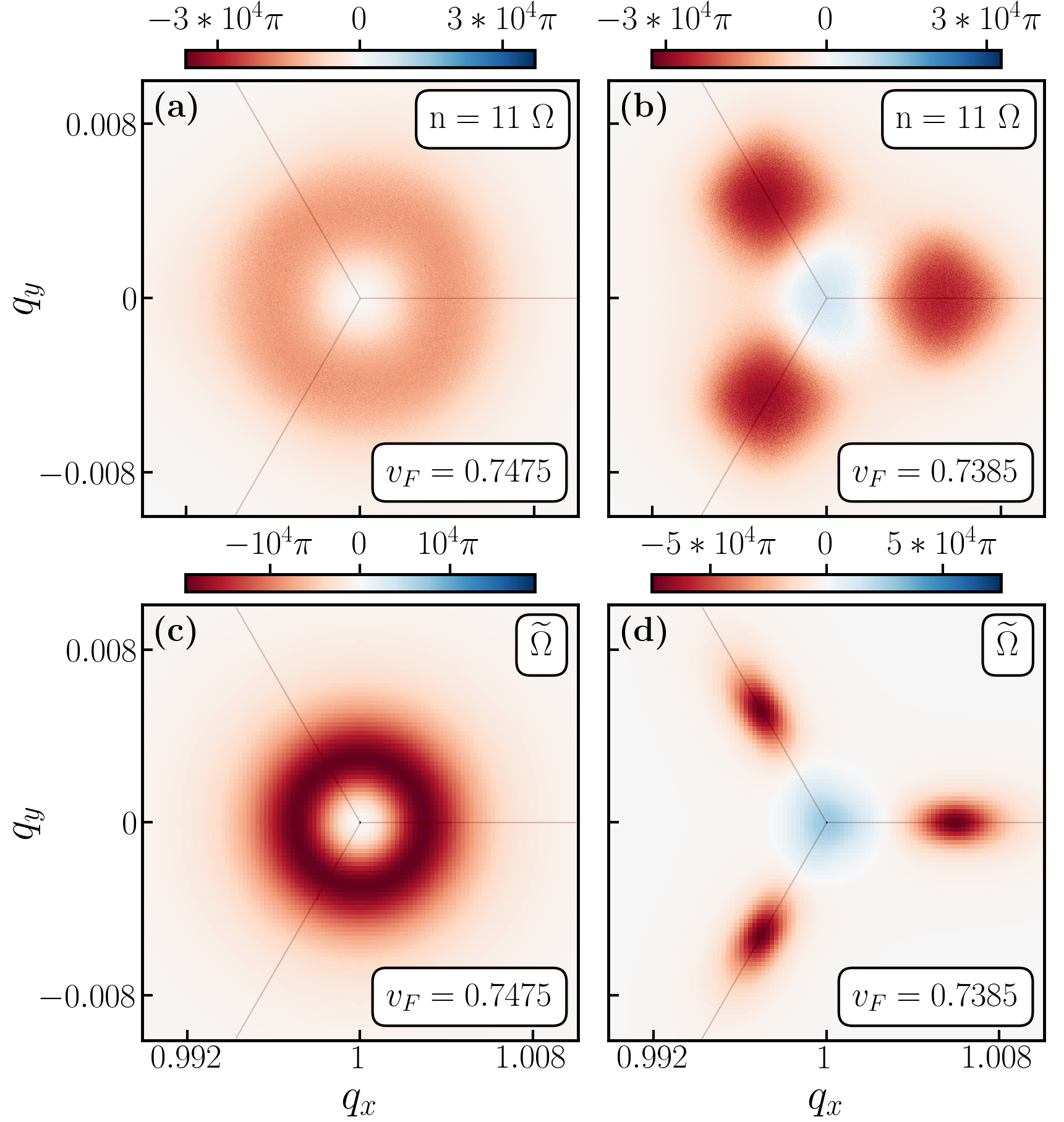}
    \caption{
    Berry curvature distributions for spinorless 3-band model and $n = 11$-dispersion model. Here $|\v{\kappa}| = 1, m = 10$, maroon lines indicate BZ boundaries, and $\lambda = 0.9999224 \neq \lambda_\mathrm{crit}^*$. (a,b) shows n-dispersion Berry curvature for $v_F = v_F^* = 0.7475$ and $v_F= 0.7385$ respectively. (c,d) shows spinorless Berry curvature for $v_F = v_F^* = 0.7475$ and $v_F = 0.7385$ respectively.
    }
    \label{fig:QBT_berry_n_disp}
\end{figure}

\begin{figure}
    \centering
    \includegraphics[width=\linewidth]{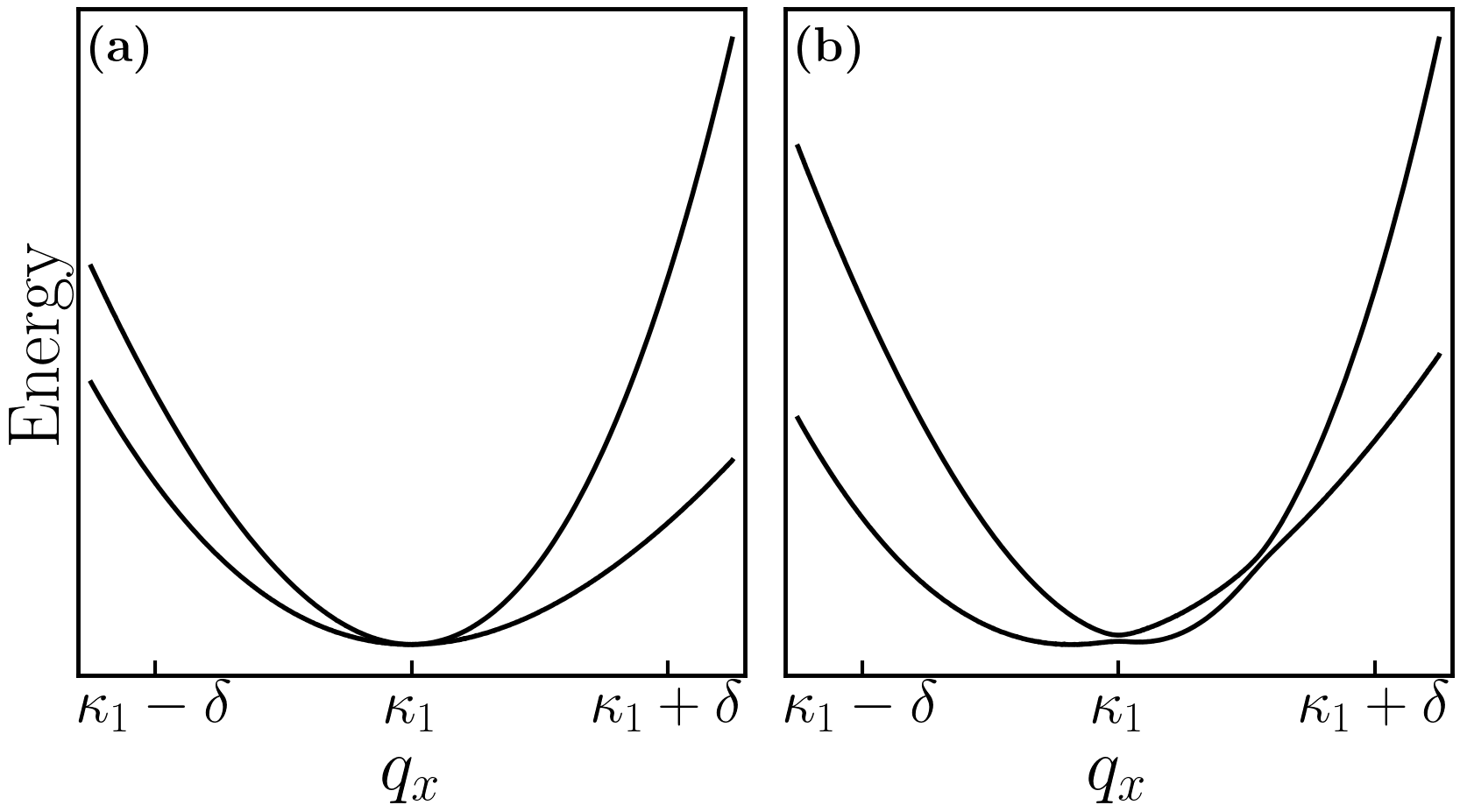}
    \caption{Energy of 2 lowest-lying bands of $n=11$-dispersion. Here $|\delta| = 0.05 |\v{\kappa}|, m = 10$.(a) shows QBT for $\lambda = \lambda_\mathrm{crit}^* = 0.9998924$ and $v_F = v_F^* = 0.7475$. (b) shows avoided crossing for $v_F = 0.6775$ and $\lambda = 0.9999724$.
    }
    \label{fig:n_disp_bands}
\end{figure}

Exactly at $\lambda = \lambda_\mathrm{crit}$, Eq.~\eqref{eq:delta_alpha} gives $\arg(\Delta) = -2\pi/3$ and $\arg(\alpha) = \pi/3$. Setting $v_F = 2|\alpha| = 3/4$ endows the three-band model with coexisting $U(1)$ and $PK$ symmetries  (see  Table~\ref{tab:emergence_crit} and App.~\ref{app:QBT_lambda}) as well as a quadratic band touching at $(q_x, q_y) = (0, 0)$.
The critical $v_F^*$ in the un-truncated model is $O(10^{-3})$ smaller than $3/4$ due to the band warping by higher order momentum-dependent terms in the potential matrix elements (Eq.~\eqref{eq:V_mat_elems}). By fixing $\lambda = \lambda_\mathrm{crit}^* = 0.9998924/ |\v{\kappa}| \approx 1/|\v{\kappa}|$ and $v_F = v_F^* =  0.7475$, we observe a QBT in Fig.~\ref{fig:n_disp_bands}(a). Gapping out the bands by perturbing $\lambda$ produces nearly identical annuli of n-dispersion and spinorless Berry curvature in Fig.~\ref{fig:QBT_berry_n_disp}(a,c) respectively.

The 3-band model predicts that perturbing $v_F$ away from $v_F^*$ breaks the $U(1)$ symmetry down to $C_3$ and splits the QBT into 4 Dirac cones (Sec.~\ref{sec:topo_dirac_cones}). We observe this QBT splitting in the n-dispersion band structure, as shown in Fig.~\ref{fig:n_disp_bands}(b). The resultant hotspots of Berry curvature are consistent in sign and distribution between n-dispersion (Fig.~\ref{fig:QBT_berry_n_disp}(b)) and spinorless Berry curvatures (Fig.~\ref{fig:QBT_berry_n_disp}(d)).

\section{Discussion}
\label{sec:discussion}

The three-band model successfully captures Berry curvature near $\v{\kappa}, \v{\kappa}'$ in parameter regimes close to critical points. As a result, the spinorless Berry curvature extensively studied in Sec.~\ref{sec:lim_topo_flat_bands} and Sec.~\ref{sec:lim_triv_flat_bands} closely approximates the near-critical Berry curvature in low-lying bands of the un-truncated model. While our un-truncated model was restricted to two components, it can be easily generalized to have more components~\cite{trithep_ahc, lambda_N_jellium}, which enables us to explore a larger region of the $(\Delta, \alpha)$ phase diagram.

Our treatment focuses on $C_3$ symmetric points, but $C_2$ symmetry also plays an important role in controlling the topology of the state (Fig.~\ref{fig:C2_ang_mom}), and our approach can be extended to cover the $\v{M}$ point of the BZ as well.
Such an extended model provides an avenue to capture the Berry curvature of both $C_3$ and $C_2$ symmetric topological crystals at the mean-field level.

The features uncovered here, in particular the topological flat bands, make the
three-band model worth studying in its own right. Because these bands are
simultaneously flat and topologically nontrivial, the model's interacting phase
diagram is likely to be especially rich, and a full many-body treatment is a
natural direction for future work. Guided by these results, an experimental
realization would then offer a versatile platform for accessing the topological
phases the model supports.

\section{Acknowledgments}

We thank Ashvin Vishwanath for discussions. A.V. acknowledges support from the Caltech Summer Undergraduate Research Fellowship.
This research is funded in part by the
Gordon and Betty Moore Foundation’s EPiQS Initiative,
Grant GBMF8683 to T.S.
The Flatiron Institute is a division of the Simons Foundation.

\bibliography{references}

\onecolumngrid
\appendix

\newpage

\section{Derivation of spectrum of effective Hamiltonian}
\label{app:effective_ham_spec_deriv}

In Sec.~\ref{sec:spectrum} we gave the eigenvalues of the 3-band Hamiltonian, so we now provide a calculation of the eigenvectors of $H$. Since $H$ is circulant at $\v{q} =0$, we know that $\epsilon_\ell(\v{q}= 0) = 2 \text{Re} (\omega^\ell \Delta)$ and $\v{u}_\ell(\v{q}=0) = \frac{1}{\sqrt{3}}(1, \omega^\ell, \overline{\omega}^\ell)^T$. From here on out, we denote the eigenvectors of $H$ as $\v{u}_\ell = (u^1_\ell, u^3_\ell, u^5_\ell)$. We have the following system of eigenvector/value equations:
\begin{equation}
\label{eq:ev_eqs}
\begin{split}
    \epsilon_\ell u^1_\ell &= u^1_\ell T_1 + u^3_\ell W_{1} + u^5_\ell \overline{W}_{5} \\
    \epsilon_\ell u^3_\ell &= u^1_\ell \overline{W}_{1} + u^3_\ell T_3 + u^5_\ell W_{3} \\
    \epsilon_\ell u^5_\ell &= u^1_\ell W_{5} + u^3_\ell \overline{W}_{3} + u^5_\ell T_5.
\end{split}
\end{equation}
We can then solve the eigenvector equations (see Eq.~\eqref{eq:ev_eqs}) of $H$ to determine $\v{u}_\ell$. For ease of computation, we fix the gauge $\text{arg}(u^1_\ell) = 0$. We give the normalized eigenvectors of $H$ in Eq.~\eqref{eq:H_anal_ev}:
\begin{equation}
    \v{u}_\ell = \frac{1}{\sqrt{N_\ell}}
    \begin{pmatrix}
        (\epsilon_\ell - T_5)(\epsilon_\ell - T_3) - |W_{3}|^2 \\
        \overline{W}_{1} (\epsilon_\ell - T_5) + W_{3} W_{5} \\
        W_{5} (\epsilon_\ell - T_3) + \overline{W}_{1} \overline{W}_{3}
    \end{pmatrix}.
    \label{eq:H_anal_ev}
\end{equation}
We leave the above eigenvectors in terms of the eigenvalues $\epsilon_\ell$, the energetic matrix elements $T_j, W_{m}$, and the normalization $N_\ell$ (Eq.~\eqref{eq:normalization}) for the sake of simplicity.
\begin{equation}
    \begin{split}
        N_\ell &= \bigg[(\epsilon_\ell - T_5) (\epsilon_\ell - T_3) - |W_{3}|^2 \bigg]^2 + |W_{3}|^2 (|W_{1}|^2 + |W_{5}|^2) + 2 \text{Re}(W_{1} W_{3} W_{5}) [2 \epsilon_\ell - T_5 - T_3] + |W_{1}|^2 (\epsilon_\ell - T_5)^2 \\&+ |W_{5}|^2 (\epsilon_\ell - T_3)^2.
    \end{split}
    \label{eq:normalization}
\end{equation}
Thus we have obtained exact, closed-form expressions for the eigenvectors and eigenvalues of $H$.

\section{Derivation of spinorless Berry curvature}
\label{app:spinorless_bc_deriv}

In this section, we provide a complete derivation of the spinorless Berry curvature of the bands of $H$ at the origin. To this end, we need to determine the momentum derivatives $\frac{\partial}{\partial q_x}, \frac{\partial}{\partial q_y}$ of the eigenvectors of $H$ at the origin. Recall that we express the normalized eigenvectors of $H$ in terms of the variables $W_{m}$ (Eq.~\eqref{eq:holo_ff}) and $T_j$ (Eq.~\eqref{eq:T_n_def}); we can compute the gradients of these scalars as
\begin{equation}
    \begin{split}
        \v{\nabla} W_{1} &= -\frac{\alpha}{2} \hat{q}_x - \alpha \frac{\sqrt{3}}{2} \hat{q}_y, \ \v{\nabla} W_{3} = \alpha \hat{q}_x, \ \v{\nabla} W_{5} = \frac{\alpha}{2} \hat{q}_x + \alpha \frac{\sqrt{3}}{2} \hat{q}_y \\
        \v{\nabla} T_1 &= v_F \hat{q}_x, \ \v{\nabla} T_3 = -\frac{v_F}{2} \hat{q}_x + v_F \frac{\sqrt{3}}{2} \hat{q}_y, \ \v{\nabla} T_5 = -\frac{v_F}{2} \hat{q}_x - v_F \frac{\sqrt{3}}{2} \hat{q}_y.
    \end{split}
\end{equation}
To determine the derivatives of the eigenvectors, we must first study the normalization $N$ (given in Eq.~\eqref{eq:normalization}) at the origin. We find that $N |_{\v{q} = 0}$ is given by
    \begin{equation}
    \begin{split}
        N |_{\v{q} = 0} = 16 \text{Re}(\omega^\ell \Delta)^4 + 3 |\Delta|^4 + 8 \text{Re}(\Delta^3) \text{Re}(\omega^\ell \Delta).
    \end{split}
\end{equation}
Since $\epsilon_\ell$ is a $C_3$ invariant scalar, it must be the case that $\vec{\nabla} \epsilon_\ell |_{\v{q}=0} = 0$. This is a consequence of the fact that there cannot be a smooth, $C_3$-invariant linear function that is non-zero at the origin. Thus the derivatives of $N$ are
\begin{equation}
\label{eq:norm_derivs}
    \begin{split}
        \frac{\partial N}{\partial q_x} |_{\v{q} = 0} &= 16 v_F \text{Re}(\omega^\ell \Delta)^3 - 24 \text{Re}(\overline{\Delta} \alpha) \text{Re}(\omega^\ell \Delta)^2 + 6 |\Delta|^2 \text{Re}(\overline{\Delta} \alpha) + 2 v_F \text{Re}(\Delta^3) \\
        \frac{\partial N}{\partial q_y} |_{\v{q} = 0} &= 0.
    \end{split}
\end{equation}
Some algebra gives the 6 momentum derivatives of the entries of $\v{u}_\ell$ as
\begin{equation}
\label{eq:ev_derivs}
\begin{split}
    \frac{\partial u^1_\ell}{\partial q_x} &= -\frac{1}{2 \sqrt{N}} \bigg[ \frac{1}{N} \bigg(16 v_F \text{Re}(\omega^\ell \Delta)^3 - 24 \text{Re}(\overline{\Delta} \alpha) \text{Re}(\omega^\ell \Delta)^2 + 6 |\Delta|^2 \text{Re}(\overline{\Delta} \alpha) + 2 v_F \text{Re}(\Delta^3) \bigg) \bigg(4 \text{Re}(\omega^\ell \Delta)^2 - |\Delta|^2 \bigg) \\ 
    &- 4 v_F \text{Re}(\omega^\ell \Delta) + 4 \text{Re} (\overline{\Delta} \alpha) \bigg] \\
    \frac{\partial u^1_\ell}{\partial q_y} &= 0 \\
    \frac{\partial u^3_\ell}{\partial q_x} &= -\frac{1}{2 \sqrt{N}} \bigg[ \frac{1}{N} \bigg(16 v_F \text{Re}(\omega^\ell \Delta)^3 - 24 \text{Re}(\overline{\Delta} \alpha) \text{Re}(\omega^\ell \Delta)^2 + 6 |\Delta|^2 \text{Re}(\overline{\Delta} \alpha) + 2 v_F \text{Re}(\Delta^3) \bigg) \bigg(2 \overline{\Delta} \text{Re}(\omega^\ell \Delta) + \Delta^2 \bigg)  \\
    &+ 2 \overline{\alpha} \text{Re} (\omega^\ell \Delta) - v_F \overline{\Delta} - \alpha \Delta \bigg] \\
    \frac{\partial u^3_\ell}{\partial q_y} &= \frac{1}{2 \sqrt{N}} \bigg[ -2 \sqrt{3} \overline{\alpha} \text{Re}(\omega^\ell \Delta) + \sqrt{3} \overline{\Delta} v_F + \sqrt{3} \Delta \alpha \bigg] \\
    \frac{\partial u^5_\ell}{\partial q_x} &= -\frac{1}{2 \sqrt{N}} \bigg[ \frac{1}{N} \bigg(16 v_F \text{Re}(\omega^\ell \Delta)^3 - 24 \text{Re}(\overline{\Delta} \alpha) \text{Re}(\omega^\ell \Delta)^2 + 6 |\Delta|^2 \text{Re}(\overline{\Delta} \alpha) + 2 v_F \text{Re}(\Delta^3) \bigg) \bigg(2 \Delta \text{Re}(\omega^\ell \Delta) + \overline{\Delta}^2 \bigg)  \\
    &+ 2 \alpha \text{Re}(\omega^\ell \Delta) - \Delta v_F - \overline{\Delta} \overline{\alpha} \bigg] \\
    \frac{\partial u^5_\ell}{\partial q_y} &= \frac{1}{2 \sqrt{N}} \bigg[2 \sqrt{3} \alpha \text{Re}(\omega^\ell \Delta) - \sqrt{3} \Delta v_F - \sqrt{3} \overline{\Delta} \overline{\alpha} \bigg].
\end{split}
\end{equation}
Using the above derivatives, we determine the Berry curvature corresponding to the 3 bands of $H$ at the origin to be
\begin{equation}
    \begin{split}
        \widetilde{\Omega}_\ell (0, 0) &= \frac{2 \sqrt{3}}{N^2} \bigg(8 v_F \text{Re}(\omega^\ell \Delta)^3 - 12 \text{Re}(\overline{\Delta} \alpha) \text{Re}(\omega^\ell \Delta)^2 + 3 |\Delta|^2 \text{Re}(\overline{\Delta} \alpha) + v_F \text{Re}(\Delta^3) \bigg) \\
        &* \bigg(4 \text{Re}(\omega^\ell \Delta)^2 \text{Im}(\overline{\Delta} \alpha) + 4 \text{Re}(\omega^\ell \Delta) \text{Im}(\alpha \Delta^2) - v_F \text{Im}(\Delta^3) + |\Delta|^2 \text{Im}(\overline{\Delta} \alpha)  \bigg).
    \end{split}
    \label{eq:og_berry_curvatures}
\end{equation}

\subsection{Roots of spinorless Berry curvature}
\label{app:spinorless_bc_roots_deriv}

In this section, we determine both branches of roots of $\widetilde{\Omega}_{\ell=0}(0, 0)$ for all bands. The second branch of roots of $\widetilde{\Omega}_{\ell=0}(0, 0)$ will inspire an ansatz taken (and verified) for the roots of $\widetilde{\Omega}_{\ell=\pm 1}(0, 0)$. First we parameterize $\Delta, \alpha$ in terms of magnitudes $d, a$ and angles $\phi, \theta$ as follows:
\begin{equation}
\label{eq:delta_alpha_parameterization}
    \begin{split}
        \Delta &= d e^{i \phi} \\
        \alpha &= a e^{i \theta}.
    \end{split}
\end{equation}
First we solve for the 1st branch of universal roots amongst all three bands. We use $(\phi_\ell, \theta_\ell)$ to denote the solutions to $\widetilde{\Omega}_\ell(0, 0) = 0$. Substituting the parameterization for $\alpha, \Delta$ into Eq.~\eqref{eq:og_berry_curvatures} yields
\begin{equation}
\label{eq:og_bc_app_1}
    \begin{split}
        \widetilde{\Omega}_\ell(0, 0) &= \frac{2 \sqrt{3} d^6}{N^2} \bigg(8 v_F \cos^3(\phi + \frac{2 \pi \ell}{3}) - 12 a \cos(\theta - \phi) \cos^2(\phi + \frac{2 \pi \ell}{3}) + 3 a \cos(\theta - \phi) + v_F \cos(3 \phi) \bigg) \\
        &* \bigg( 4 a \cos^2(\phi + \frac{2 \pi \ell}{3} ) \sin(\theta - \phi) + 4 a \cos(\phi + \frac{2 \pi \ell}{3}) \sin(\theta + 2 \phi) - v_F \sin(3 \phi) + a \sin(\theta - \phi) \bigg).
    \end{split}
\end{equation}
The apparent scaling of $\widetilde{\Omega}_\ell(0, 0)$ as $d^6$ might seem concerning; however, recall that $N \propto d^4$, so in reality $\widetilde{\Omega}_\ell(0, 0) \propto d^{-2}$. Roots of this expression $\phi, \theta$ will then make one of the 2 factors in the product shown in Eq.~\eqref{eq:og_bc_app_1} vanish. We study the case when the first factor vanishes. 
\begin{equation}
\label{eq:root_branch_1_eq}
    \begin{split}
        0 &= 8 v_F \cos^3(\phi + \frac{2 \pi \ell}{3}) - 12 a \cos(\theta - \phi) \cos^2(\phi + \frac{2 \pi \ell}{3}) + 3 a \cos(\theta - \phi) + v_F \cos(3 \phi).
    \end{split}
\end{equation}
If $\phi$ is a solution to this equation, $\phi + \pi$ is also a solution since every term in the equation reverses sign, as $\cos(x + \pi) = -\cos(x)$ so the original equation holds (since the LHS is 0).  We find that the solution satisfies
\begin{equation}
\begin{split}
    \theta_\ell &= \phi_\ell + \cos^{-1} \bigg(  \frac{v_F}{a} \frac{[\cos(3 \phi_\ell) + 2 \cos(\phi_\ell + \frac{2 \pi \ell}{3})]}{[1 + 2 \cos(2 \phi_\ell + \frac{4 \pi \ell}{3})]}  \bigg)
\end{split}
\end{equation}
This relation simplifies to
\begin{equation}
\begin{split}
    \theta_\ell &= \phi_\ell + \cos^{-1} \bigg( \frac{v_F}{a} \cos(\phi_\ell + \frac{2 \pi}{3} \ell) \bigg).
\end{split}
\end{equation}
To obtain the second set of solutions, we first solve the case where $\ell = 0$:
\begin{equation}
    \begin{split}
        0 &= 4 a \cos^2(\phi) \sin(\theta - \phi) + 4 a \cos(\phi) \sin(\theta + 2 \phi) - v_F \sin(3 \phi) + a \sin(\theta - \phi) \\
        \implies 0 &= \bigg( 1 + 2 \cos(2 \phi) \bigg) \bigg (3 a \cos(\phi) \sin(\theta) - v_F \sin(\phi) + a \cos(\theta)\sin(\phi) \bigg).
    \end{split}
\end{equation}
We disregard the solution to $0 = 1 + 2 \cos(2 \phi)$, as $\widetilde{\Omega}_0(0, 0)$ is singular here $(\phi = \frac{2 \pi}{3}, \frac{\pi}{3})$. We analyze the other factor:
\begin{equation}
    \begin{split}
        0 &= 3 a \cos(\phi) \sin(\theta) - v_F \sin(\phi) + a \cos(\theta)\sin(\phi) \\
        &= 2 \sin(\theta + \phi) + \sin(\theta - \phi) - \frac{v_F}{a} \sin(\phi) \\
        &= 3 \sin(\theta) \cos(\phi) + \cos(\theta) \sin(\phi) - \frac{v_F}{a} \sin(\phi) \\
        \implies \phi &= \tan^{-1} \bigg( \frac{3 \sin(\theta)}{\frac{v_F}{a} - \cos(\theta)} \bigg).
    \end{split}
\end{equation}
Now that we have obtained this branch of solutions for $\ell = 0$, determination of the corresponding roots for $\ell = -1, 1$ follows in a straightforward manner. The following equation gives the root of $\tilde \Omega$ for  $\ell \in \{-1, 0, 1\}$:
\begin{equation}
\begin{split}
    \frac{v_F}{a} \sin(3 \phi) &= 4 \cos^2(\phi + \frac{2 \pi \ell}{3}) \sin(\theta - \phi) + 4 \cos(\phi + \frac{2 \pi \ell}{3}) \sin(\theta + 2 \phi) + \sin(\theta - \phi).
\end{split}
\end{equation}
Let $\theta_0, \phi_0$ denote the solution $\widetilde{\Omega}_0(0, 0) = 0$ corresponding to the 2nd branch. These solutions satisfy the equation
\begin{equation}
    \frac{v_F}{a} \sin(3 \phi_0) = 4 \cos^2(\phi_0 + 0*\frac{2 \pi}{3}) \sin(\theta_0 - \phi_0) + 4 \cos(\phi_0 + 0*\frac{2 \pi}{3}) \sin(\theta_0 + 2 \phi_0) + \sin(\theta_0 - \phi_0) .
\end{equation}
From this it follows that the roots for $\ell = -1$ and $\ell = 1$ are $(\theta_1, \phi_1) = (\theta_0 + \frac{4 \pi}{3}, \phi_0 + \frac{\pi}{3})$ and $(\theta_{-1}, \phi_{-1}) = (\theta_0 + \frac{2 \pi}{3}, \phi_0 + \frac{2 \pi}{3})$ respectively.

We can now describe the branches of roots universal across all 3 bands. The 1st branch of roots is given as
\begin{equation}
    \begin{split}
        \theta_\ell &= \phi_\ell + \cos^{-1} \bigg( \frac{v_F}{a} \cos(\phi_\ell + \frac{2 \pi}{3} \ell) \bigg).
    \end{split}
\end{equation}
The second branch of roots is given as 
\begin{equation}
    \begin{split}
        \phi_\ell &= n \pi + \tan^{-1} \bigg(\frac{3 \sin(\theta_\ell + \frac{2\pi \ell}{3})}{\frac{v_F}{a} - \cos(\theta_\ell + \frac{2\pi \ell}{3})} \bigg) + \frac{\pi}{3} \ell .
    \end{split}
\end{equation}
for $n \in \mathbb {Z}$.

\section{Expansion of Hamiltonian eigenvectors}
\label{app:berry_scales}

In this section, we derive the large-$|q|$ expansion of the eigenvectors of $H$ quoted in Eq.~\eqref{eq:expanded_u}. The closed-form eigenvector in Eq.~\eqref{eq:H_anal_ev} is exact for \emph{any} eigenvalue of $H$, so the expansion below applies equally to each of the three bands; comparing Eq.~\eqref{eq:H_anal_ev} to Eq.~\eqref{eq:expanded_u}, only the topmost entry is approximated, and we need only derive that expansion. We index the three eigenvalues by $s \in \{-1,0,1\}$ as in Eq.~\eqref{eq:eigenvals} and expand to second order in $\eta = |\alpha|/v_F$ and $\rho = |\Delta|/(v_F |q|)$. Which $\v{q}=0$ angular-momentum band (Sec.~\ref{sec:spectrum}) a given index $s$ continues to is $\arg(\Delta)$-dependent, but this assignment plays no role in what follows: the construction is identical for all three roots.

The eigenvalues are the roots of the depressed cubic $\epsilon_s^{\,3} - B\,\epsilon_s - C = 0$, with $B, C$ defined in Eq.~\eqref{eq:B_C_defs} as
\begin{equation}
    \begin{split}
        B &= \frac{3}{2}\!\left(|\alpha|^2 + \frac{v_F^2}{2}\right)(q_x^2 + q_y^2) + 3 |\Delta|^2, \\
        C &= \frac{1}{4} q_x (q_x^2 - 3 q_y^2)\!\left(v_F^3 - 3 v_F |\alpha|^2 + 2\,\mathrm{Re}[\alpha^3]\right) - \frac{3}{2}(q_x^2 + q_y^2)\,\mathrm{Re}\!\left(\Delta \alpha^2 + 2 v_F \alpha \overline{\Delta}\right) + 2\,\mathrm{Re}(\Delta^3) .
    \end{split}
\end{equation}
With $\phi = \arg(\Delta) - \arg(\alpha)$ and $\theta = \arg(q)$, substituting $|\alpha| = v_F \eta$ and $|\Delta| = v_F |q| \rho$ into $B$ gives the exact identity
\begin{equation}
    B = v_F^2 |q|^2 \left(\frac{3}{4} + \frac{3}{2}\eta^2 + 3\rho^2\right),
\end{equation}
which we split into a leading piece and a second-order correction,
\begin{equation}
    B = B_0 + B_2,\qquad B_0 = \frac{3}{4}\, v_F^2 |q|^2,\qquad B_2 = v_F^2 |q|^2\!\left(\frac{3}{2}\eta^2 + 3\rho^2\right).
\end{equation}
Expanding $C$ to second order in $\eta,\rho$ likewise yields $C \approx C_0 + C_2$ with
\begin{equation}
    C_0 = \frac{1}{4} v_F^3 |q|^3 \cos(3\theta),\qquad
    C_2 = v_F^3 |q|^3 \left(-\frac{3}{4}\eta^2 \cos(3\theta) - 3\,\eta\rho\cos\phi\right),
\end{equation}
where the $\eta^3$, $\eta^2 \rho$, and $\rho^3$ contributions to $C$ have been dropped.

At zeroth order in $\eta,\rho$, the cubic reduces to $x_s^{\,3} - B_0\, x_s - C_0 = 0$, i.e.\ $4 t_s^{\,3} - 3 t_s - \cos(3\theta) = 0$ for $t_s = x_s / (v_F |q|)$. Using the triple-angle identity $4\cos^3\psi - 3\cos\psi = \cos(3\psi)$, the three real roots are
\begin{equation}
    x_s = v_F |q|\, t_s,\qquad t_s \equiv \cos\!\left(\theta - \frac{2\pi s}{3}\right),\qquad s \in \{-1,0,1\}.
    \label{eq:x_ell}
\end{equation}
Each leading-order term $v_F |q|\, t_s$ coincides with one of the kinetic energies of Eq.~\eqref{eq:T_n_def}, namely $x_s = T_{j(s)}$ with $j(s) \in \{1,3,5\}$ given by $j \bmod 3 \equiv (1-s)$, i.e.\ $j(0)=1$, $j(1)=3$, $j(-1)=5$.

Now write $\epsilon_s = x_s + \delta_s$ and expand $\epsilon_s^{\,3} - B\,\epsilon_s - C = 0$ to second order in $\eta,\rho$ (and to first order in $\delta_s$). The unperturbed identity $x_s^{\,3} - B_0\, x_s - C_0 = 0$ cancels the zeroth-order terms and leaves
\begin{equation}
    \delta_s = \frac{B_2\, x_s + C_2}{3 x_s^{\,2} - B_0}.
\end{equation}
Using $x_s = v_F |q|\, t_s$, the denominator simplifies to $3 x_s^{\,2} - B_0 = \tfrac{3}{4}\, v_F^2 |q|^2 (4 t_s^{\,2} - 1)$, and substituting $B_2$ and $C_2$ gives
\begin{equation}
    \epsilon_s \approx x_s + f_s,\qquad
    f_s = \frac{v_F |q|}{4 t_s^{\,2} - 1}\Big[\!\left(2 t_s - \cos(3\theta)\right) \eta^2 - 4\cos(\phi)\,\eta\rho + 4 t_s\,\rho^2\Big].
    \label{eq:f_ell}
\end{equation}
The denominator $4 t_s^{\,2} - 1$ vanishes precisely when the unperturbed cubic has a double root, at $\sin(3\theta) = 0$; near these points the non-degenerate expansion breaks down and a degenerate treatment is required.

Finally, expanding $(\epsilon_s - T_5)(\epsilon_s - T_3)$ from Eq.~\eqref{eq:H_anal_ev} with $\epsilon_s \approx T_{j(s)} + f_s$ and dropping the fourth-order term $f_s^{\,2}$,
\begin{equation}
    \begin{split}
        (\epsilon_s - T_5)(\epsilon_s - T_3) &\approx \big(T_{j(s)} - T_3\big)\big(T_{j(s)} - T_5\big) + f_s\,\big(2 T_{j(s)} - T_3 - T_5\big).
    \end{split}
\end{equation}
The remaining entries of Eq.~\eqref{eq:expanded_u} are taken from Eq.~\eqref{eq:H_anal_ev} without approximation, so this recovers the eigenvector of the band dominated by $T_{j(s)}$. Since the construction is identical for all three roots, it yields the eigenvector of each band; for the band carrying $\v{q}=0$ angular momentum $\ell$, it gives $\v{u}_\ell$ as quoted in Eq.~\eqref{eq:expanded_u}.

\section{Spectra of emergent Hamiltonians}

In this section we extract the eigenvalues and eigenvectors of several of the Hamiltonians mentioned in Sec.~\ref{sec:lim_topo_flat_bands} and Sec.~\ref{sec:lim_triv_flat_bands}. 

\subsection{$U(1)$ symmetric Hamiltonian}
\label{app:U(1)}

Consider $H_\omega$ in Eq.~\eqref{eq:U(1)_H}. We project $H_\omega$ to a subspace orthonormal to $\v{u}_0$ (see Eq.~\eqref{eq:U(1)_ev0}) using the vectors $\v{v}_- = (0, 1, 0)^T, \v{v}_+ = \frac{1}{|q| \sqrt{P^2 + M^2}} (M \overline{q}, 0,  P q)^T$. The reduced Hamiltonian becomes
\begin{equation}
    H_2 = \frac{(\Delta_3 + \widetilde{\Delta})}{2} I + \frac{(\Delta_3 - \widetilde{\Delta})}{2} \sigma^z + \sqrt{M^2 + P^2} |q| \sigma^x .
\end{equation}
We can extract the eigenvalues and eigenvectors in the usual way:
\begin{equation}
    \epsilon_\pm = \frac{(\Delta_3 + \widetilde{\Delta})}{2} \pm \frac{1}{2} \sqrt{(\Delta_3 - \widetilde{\Delta})^2 + 4 (M^2 + P^2)|q|^2}
\end{equation}
and
\begin{equation}
\label{eq:U(1)_PK_ev_pm}
\begin{split}
    \v{u}_- &=  -\sqrt{M^2 + P^2} |q| \v{v}_- + (\epsilon_+ - \widetilde{\Delta}) \v{v}_+ \\
    \v{u}_+ &= (\epsilon_+ - \widetilde{\Delta})\v{v}_- + \sqrt{M^2 + P^2} |q| \v{v}_+ .
\end{split}
\end{equation}

\subsection{$C_3, PK$ symmetric Hamiltonian}
\label{app:C3_PK}

Consider $H_\omega$ in Eq.~\eqref{eq:gapless_dirac_cones_H} along the line $q_y = 0$, and recall this Hamiltonian commutes with $P_+$. The spectrum of $P_+$ can be identified as
\begin{equation}
\label{eq:Perm_ev}
    \begin{split}
        \lambda &= 1, \v{v}_0 = (0, 1, 0)^T \\
        \lambda_\pm &= \pm 1, \v{v}_\pm = (1, 0, \pm 1)^T .
    \end{split}
\end{equation}
Expanding $H_\omega$ in the orthonormal basis spanned by $\v{v}_0, \v{v}_+$ yields
\begin{equation}
    H_2 = \frac{(\Delta_3 + \widetilde{\Delta} + M q_x)}{2} I + \frac{(\Delta_3 - \widetilde{\Delta} - M q_x)}{2} \sigma^z + J \sqrt{2} q_x \sigma^x .
\end{equation}
From this Hamiltonian, we determine the other 2 eigenvalues of $H_\omega$ to be
\begin{equation}
\label{eq:C3_PK_eigenval_pm}
    \epsilon_\pm = \frac{(\Delta_3 + \widetilde{\Delta} + M q_x)}{2} \pm \frac{1}{2} \sqrt{(\Delta_3 - \widetilde{\Delta} - M q_x)^2 + 8 J^2 q_x^2 } .
\end{equation}
The corresponding eigenvectors are
\begin{equation}
    \begin{split}
        \v{u}_- &= -J \sqrt{2}\, q_x\, \v{v}_0 + (\epsilon_+ - \widetilde{\Delta} - M q_x)\, \v{v}_+ \\
        \v{u}_+ &= (\epsilon_+ - \widetilde{\Delta} - M q_x)\, \v{v}_0 + J \sqrt{2}\, q_x\, \v{v}_+ .
    \end{split}
\end{equation}

\subsection{$U(1) \times U(1)$ symmetric Hamiltonian}
\label{app:U(1)_U(1)}

Consider $H_\omega$ in Eq.~\eqref{eq:U(1)_U(1)_H}. We can project $H_\omega$ in a basis that is orthonormal to $\v{u}_0$ (see Eq.~\eqref{eq:U(1)_U(1)_ev0}) using $(1, 0, 0)^T, (0, 0, 1)^T$. This yields the following effective $2 \times 2$ Hamiltonian
\begin{equation}
    H_{2} = \frac{(\Delta_1 + \Delta_5)}{2} I + \frac{(\Delta_1 - \Delta_5)}{2} \sigma^z + J q_x \sigma^x - J q_y \sigma^y .
\end{equation}
The band energies can be computed as
\begin{equation}
    \epsilon_\pm = \frac{(\Delta_1 + \Delta_5)}{2} \pm \frac{1}{2} \sqrt{(\Delta_1 - \Delta_5)^2 + 4 J^2 (q_x^2 + q_y^2)}, 
\end{equation}
and the corresponding eigenvectors are
\begin{equation}
\label{eq:U(1)_U(1)_ev_pm}
    \begin{split}
        \v{u}_- &= (-i J q_y - J q_x, 0, \epsilon_+ - \Delta_5)^T\\
        \v{u}_+ &= (\epsilon_+ - \Delta_5, 0, J q_x - i J q_y)^T .
    \end{split}
\end{equation}

\section{Emergent symmetries and quadratic band touching from tuning $\lambda, v_F$}
\label{app:QBT_lambda}

In this section, we show that setting  $\lambda = 1 / |\v{\kappa}|, v_F = 3/4$ in Eq.~\eqref{eq:delta_alpha} results in a Hamiltonian with $U(1)$ and $PK$ symmetries. 

Evaluating Eq.~\eqref{eq:delta_alpha} at criticality, $\lambda = 1/|\v{\kappa}|$,
with $|\v{\kappa}| = 1$ and $v_1 = v_2 = -1$ yields
\begin{equation}
\label{eq:lambda_jell_crit_params}
    \Delta = \frac{\overline{\omega}}{4}, \qquad \alpha = \frac{3}{8} e^{i \pi/3},
\end{equation}
i.e., $\arg(\Delta) = -2\pi/3$ and $\arg(\alpha) = \pi/3$. From Table~\ref{tab:emergence_crit}, tuning
\begin{equation}
    v_F = -2 \, \text{Re}(\omega \alpha) = 2 |\alpha| = \frac{3}{4}
\end{equation}
sets $J_5 = 0$, producing a $U(1)$ symmetry generated by
$Q_5 = \text{diag}(1, -1, 0)$.  Furthermore, $\arg(\Delta) = -2\pi/3$ gives $\Delta_1 = \Delta_3 \equiv \widetilde{\Delta} = -|\Delta|$, and $\arg(\alpha) = \pi/3$ gives $J_1 = J_3 \equiv J$ with $J = 3|\alpha|/2$, so the $PK$ criteria of Table~\ref{tab:emergence_crit} are satisfied for
$(j, k) = (1, 3)$. The resulting Hamiltonian is
\begin{equation}
\label{eq:U(1)_PK_13_H}
    H_\omega = \begin{pmatrix}
        \widetilde{\Delta} & 0 & J q \\
        0 & \widetilde{\Delta} & J \overline{q} \\
        J \overline{q} & J q & \Delta_5
    \end{pmatrix},
\end{equation}
where $\Delta_5 = 2 |\Delta|$, and the anti-unitary symmetry is generated by
\begin{equation}
    P_{13} = \begin{pmatrix}
        0 & 1 & 0 \\
        1 & 0 & 0 \\
        0 & 0 & 1
    \end{pmatrix}.
\end{equation}
As in Sec.~\ref{sec:QBT_U(1)_PK}, the anti-unitary symmetry enforces vanishing spinorless Berry curvature across all of $\v{q}$-space.

This Hamiltonian possesses a singular topological flat band at energy $\widetilde{\Delta}$:
\begin{equation}
\label{eq:U(1)_PK_13_ev0}
    \epsilon_0 = \widetilde{\Delta}, \quad \v{u}_0 = (q, -\overline{q}, 0)^T.
\end{equation}
The two remaining bands are rotationally invariant, as required by the $U(1)$ symmetry:
\begin{equation}
\label{eq:U(1)_PK_13_ev_pm}
    \epsilon_\pm = \frac{\widetilde{\Delta} + \Delta_5}{2}
    \pm \frac{1}{2} \sqrt{(\Delta_5 - \widetilde{\Delta})^2 + 8 J^2 |q|^2}.
\end{equation}
At $\v{q} = 0$, the flat band is degenerate with the lower band $\epsilon_-$. Since $\v{u}_0$ has twice the phase winding of a Dirac-like spinor, the gapless point at $\v{q} = 0$ is a quadratic band touching between the two lowest-energy
bands. Perturbing $v_F$ away from $2|\alpha|$ breaks the $U(1)$ symmetry down to $C_3$ and splits the QBT into 4 vorticity-conserving Dirac cones, in direct analogy with Sec.~\ref{sec:topo_dirac_cones}; this is the perturbation probed numerically in Fig.~\ref{fig:QBT_berry_n_disp}(b, d).

\section{Comparison between quadratic dispersion Berry curvature and 3-band Berry curvature}
\label{app:n=2_3band_berry_comp}

\begin{figure}
    \centering
    \includegraphics[width=0.75\linewidth]{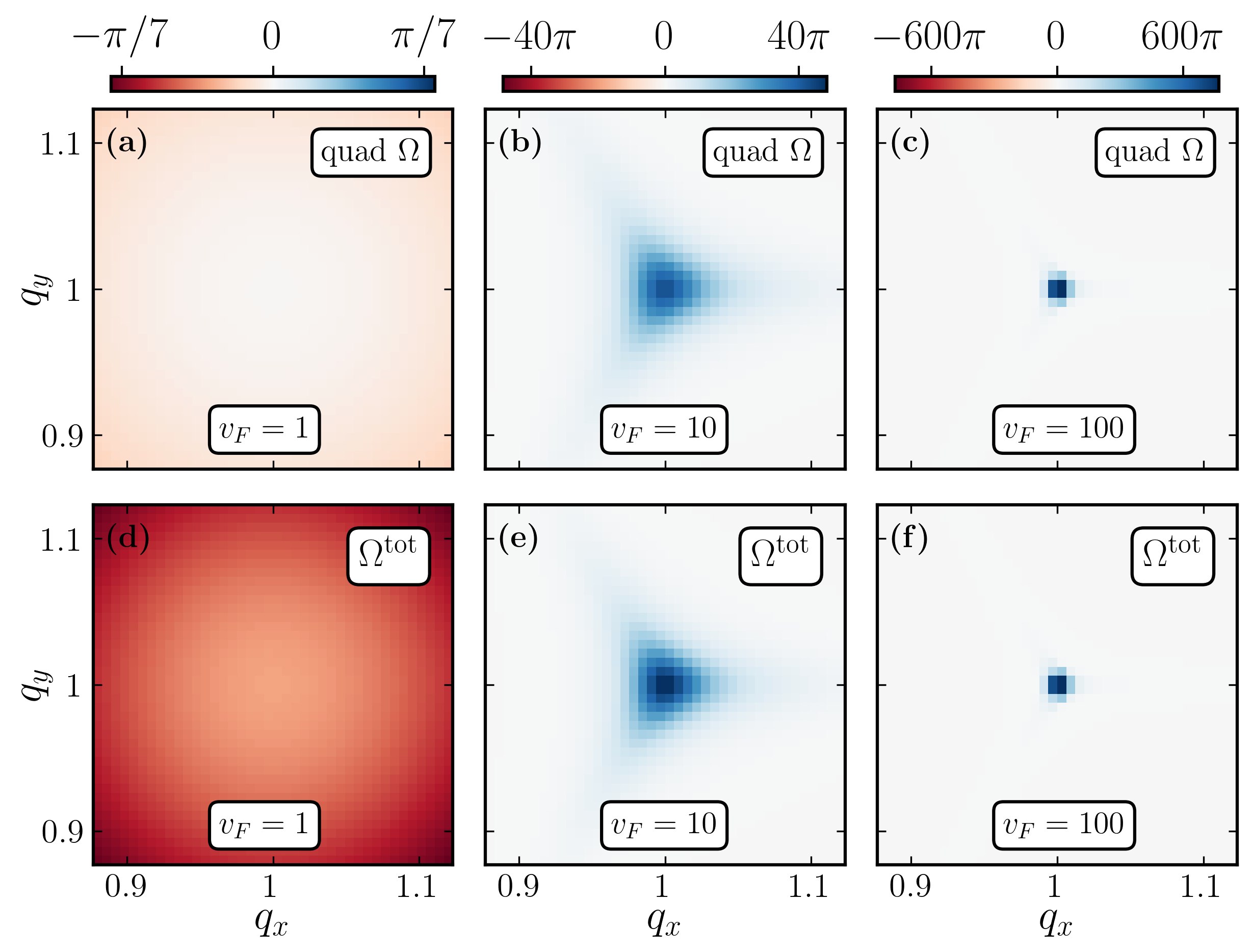}
    \caption{Comparison of Berry curvature of quadratically dispersing model (top row) and total 3-band Berry curvature (bottom). Parameters used are $v_1 = v_2 = -1, |\v{\kappa}| = 1, \lambda = 1.2 / |\v{\kappa}|$. (a-c) show quadratic dispersion Berry curvatures for $v_F = 1, 10, 100$ respectively. (d-f) show total 3-band Berry curvatures for $v_F = 1, 10, 100$ respectively.
    }
    \label{fig:mBZ_benchmarks_1p2}
\end{figure}

In this section, we provide further evidence that for the quadratic dispersion in Eq.~\eqref{eq:quad_disp}, the total 3-band Berry curvature ($\Omega^\mathrm{tot}$) and the spinorless Berry curvature ($\widetilde{\Omega}$) both closely approximate the true Berry curvature ($\mathrm{quad} \ \Omega$) for $v_F \gg |\alpha|$. In the main text, we numerically demonstrated that in the subcritical regime ($\lambda = 0.8/|\v{\kappa}|$), the total 3-band Berry curvature closely matches the true Berry curvature for $v_F \gg |\alpha|$ (see Fig.~\ref{fig:mBZ_benchmarks_0p8}). We observe a similar correspondence between the 2 Berry curvatures in the supercritical regime ($\lambda = 1.2 / |\v{\kappa}|$), as visible in Fig.~\ref{fig:mBZ_benchmarks_1p2}.

\begin{figure}
    \centering
    \includegraphics[width=0.75\linewidth]{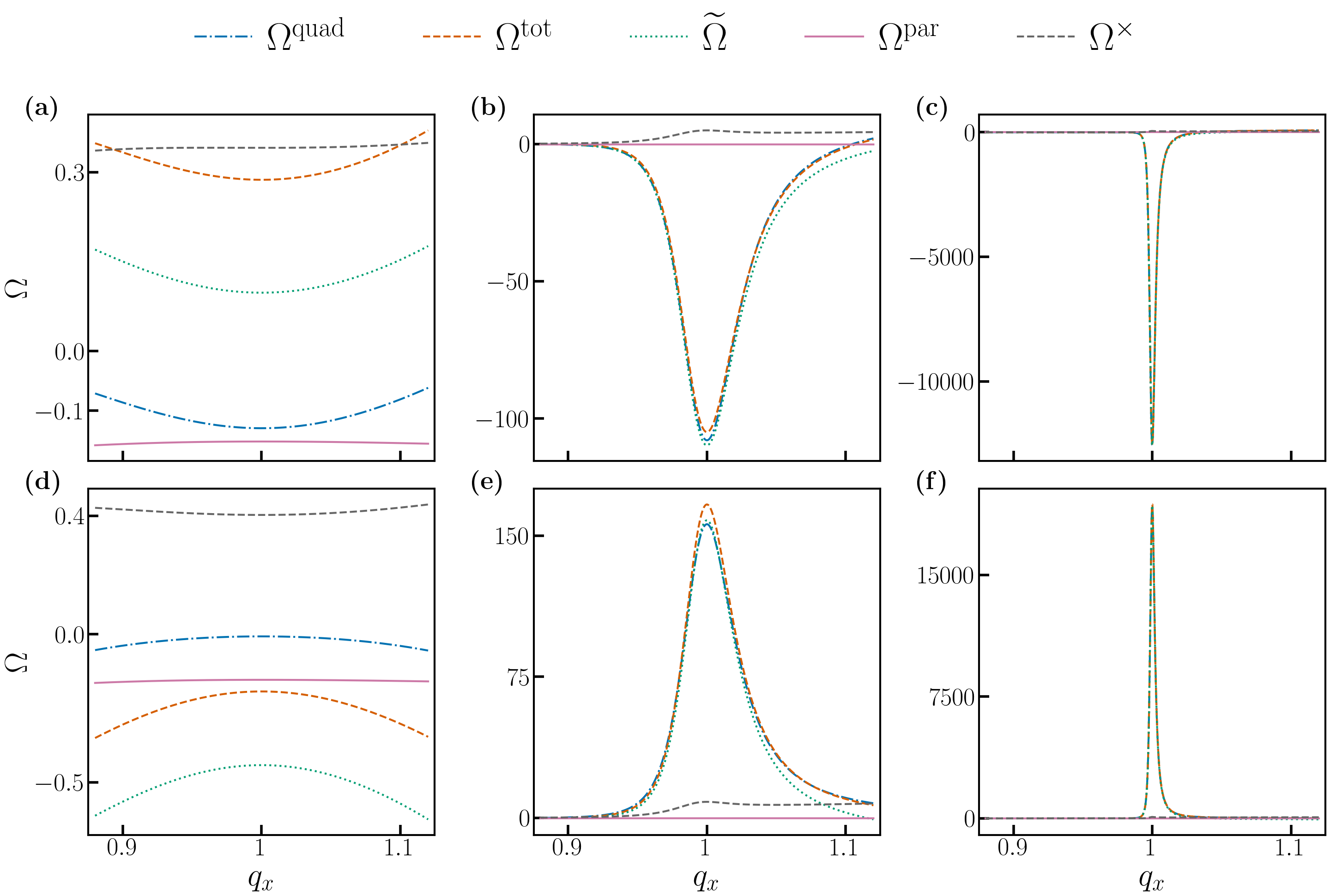}
    \caption{Comparison of Berry curvature of quadratically dispersing model ($\Omega^\mathrm{quad}$) to $\Omega^\mathrm{tot}, \widetilde{\Omega}, \Omega^\mathrm{par}, \Omega^\times$ along the cut $q_y = 0$. Parameters used are $v_1 = v_2 = -1, |\v{\kappa}| = 1$. (a-c) show Berry curvatures for $\lambda = 0.8 / |\v{\kappa}|$ and $v_F = 1, 10, 100$ respectively. (d-f) show Berry curvatures for $\lambda = 1.2 / |\v{\kappa}|$ and $v_F = 1, 10, 100$ respectively. In (c, f), the curves $\{\Omega^\mathrm{quad}, \Omega^\mathrm{tot}, \widetilde{\Omega} \}$ all overlap, and the curves $\{\Omega^\mathrm{par}, \Omega^\times\}$ overlap.}
    \label{fig:mBZ_benchmarks_component_line_cuts}
\end{figure}

We can additionally compare each of the three contributions to the total Berry curvature against the true Berry curvature. In both the subcritical regime (see (a-c) of Fig.~\ref{fig:mBZ_benchmarks_component_line_cuts}) and the supercritical regime (see (d-f) of Fig.~\ref{fig:mBZ_benchmarks_component_line_cuts}), we find that for $v_F \gg |\alpha|$, the total 3-band Berry curvature and spinorless Berry curvature capture the behavior of the true Berry curvature. The fact that the spinorless Berry curvature can closely approximate the true Berry curvature in particular regimes is a promising indication that the singular flat bands and quadratic band touchings observed in Sec.~\ref{sec:lim_topo_flat_bands} and Sec.~\ref{sec:lim_triv_flat_bands} might be experimentally accessible. As an aside, the dominance of the spinorless Berry curvature in near-critical regimes observed in Fig.~\ref{fig:mBZ_benchmarks_component_line_cuts} is consistent with the conclusions regarding Berry curvature scaling in Sec.~\ref{sec:bc_scaling}.

\section{Chern number modulo 2}
\label{app:chern_mod_2}

In this section, we discuss the inability of symmetry indicators to capture the full Chern number of a system. In particular, we focus on the n-dispersion model, noting that the $C_3$ symmetry indicator alone cannot detect changes in Chern number modulo 2. In addition, we propose a rough description of how the three-band model might be modified to capture changes in Chern modulo 6 rather than just changes modulo 3.

\begin{figure*}[h]
    \centering
    \includegraphics[width=0.75\linewidth]{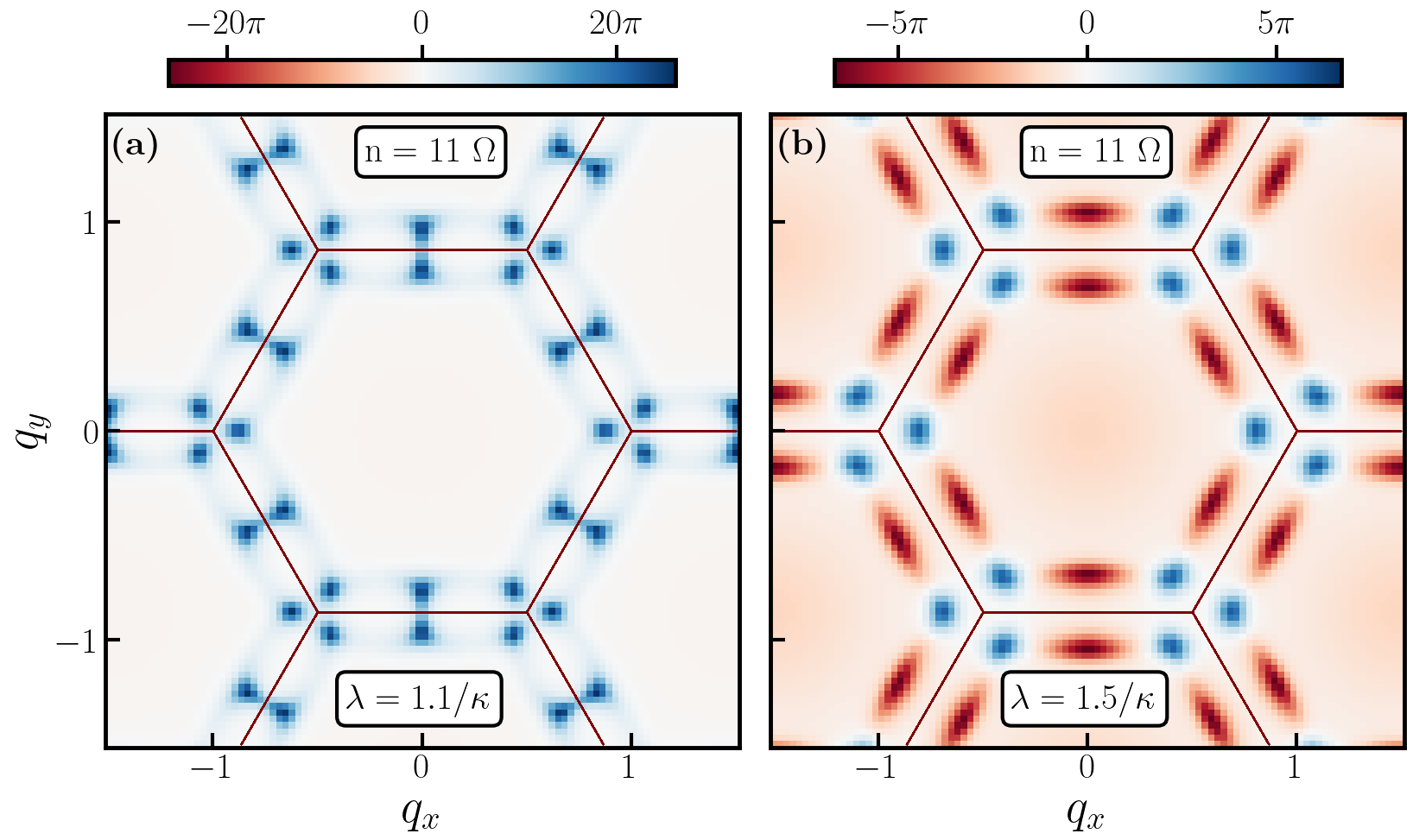}
    \caption{Berry curvature for n-dispersion model with $n = 11$ and $|\v{\kappa}| = 1$ at $v_F = 1.5$, $m = 10 |\v{\kappa}|^n$, $\hat{V} = \text{diag}(-1, -1)$. Maroon lines denote BZ boundaries. (a) shows Berry curvature for $\lambda = 1.1/ |\v{\kappa}|$. (b) shows Berry curvature for $\lambda = 1.5 /|\v{\kappa}|$.}
    \label{fig:C2_ang_mom}
\end{figure*}

As shown in Fig.~\ref{fig:C2_ang_mom}, tuning $\lambda$ past some threshold changes the sign of the Berry curvature in the vicinity of the $C_2$ invariant $M$ points of the BZ. This signature is accompanied by a change in Chern number modulo 2 due to a gap closing for some $(\lambda, v_F)_\mathrm{crit}$ at the $\v{M}$ points. However, the Chern number modulo 3 is unchanged after the closing and re-opening of a gap at $\v{M}$. This phenomenon highlights the failure of symmetry indicators to capture the full Chern number of a system.

Interestingly, note that the sign and form of the Berry curvature near the $\v{\kappa}, \v{\kappa}'$ points of the BZ at $\lambda = 1.1/|\v{\kappa}|$ and $\lambda = 1.5/\v{\kappa}$ are quite similar. This approximate invariance of Berry curvature near $\v{\kappa}, \v{\kappa}'$ suggests that a reduced model attempting to capture Berry curvature near $\v{M}, \v{\kappa}, \v{\kappa}'$ could likely be approximated by a block-diagonal Hamiltonian. Moreover, the invariance of Berry curvature at $\v{\kappa}, \v{\kappa}'$ indicates the robustness of the 3-band model for Berry curvature despite the gap closing at another point in the BZ.

\end{document}